\documentclass[a4paper, 12 pt]{article}
\usepackage[T1]{fontenc}
\usepackage[utf8]{inputenc}
\usepackage[english]{babel}
\usepackage{amsmath}
\usepackage{amssymb}
\usepackage{braket}
\usepackage{mathtools}
\usepackage{dsfont}
\usepackage{subcaption}
\usepackage{booktabs}
\usepackage{yfonts}
\usepackage{setspace}
\usepackage{cite}
\usepackage{verbatim}

\usepackage{amstext} 
\usepackage{array}   
\newcolumntype{C}{>{$}c<{$}} 

\usepackage{geometry}
\usepackage[usenames,dvipsnames]{color}
\usepackage{hyperref}
\hypersetup{
  colorlinks,
  citecolor=Blue,
  linkcolor=Blue,
  urlcolor=Blue}

\numberwithin{equation}{section}

\def\be{\begin{equation}}
\def\ee{\end{equation}}
\def\rme{{\rm e}}

\newcommand{\diff}{\mathrm{d}}
\newcommand{\ii}{\mathrm{i}} 

\def\aa{\mathsf{a}}
\def\bb{\mathsf{b}}

\def\poly{{\Psi}} 
\def\crho{{c}}
\def\mm{{\mathsf{m}}}
\def\kk{{\mathsf{k}}}
\def\mn{{\cal M}_N}
\def\ms{{\cal M}_S}

\begin{document}

\pagestyle{empty}

\begin{center}

$\,$
\vskip 0.5cm

{\Large{\bf Localization of the 5D supergravity action\\[3mm] and Euclidean saddles for the black hole index}}

\vskip 1cm

{
Davide Cassani,${}^{\textrm a}$
Alejandro Ruip\'erez,${}^{\textrm c}$
Enrico Turetta${}^{\textrm a,b}$
}

\vskip 1cm

\end{center}

\renewcommand{\thefootnote}{\arabic{footnote}}

\begin{center}
$^{\textrm a}$ {\it INFN, Sezione di Padova, Via Marzolo 8, 35131 Padova, Italy},\\ [2mm] 
$^{\textrm b}${\it Dipartimento di Fisica e Astronomia ``Galileo Galilei'', Universit\`a di Padova,\\Via Marzolo 8, 35131 Padova, Italy}\\[2mm]
$^{\textrm c}$ {\it Dipartimento di Fisica, Universit\`a di Roma ``Tor Vergata'' \& Sezione INFN Roma 2,\\ Via della Ricerca Scientifica 1, 00133, Roma, Italy}

\vskip 3cm

 {\bf Abstract} 
\end{center}

{\noindent We investigate equivariant localization of the gravitational on-shell action in odd dimensions, focusing on five-dimensional ungauged supergravity. We analyze the conditions for cancellation of boundary terms, so that the full action integral is given in terms of the odd-dimensional analog of the nuts and bolts of Gibbons-Hawking. We specialize to supersymmetric configurations with an additional ${\rm U}(1)$ symmetry preserving the supercharge and provide a formula for the localized on-shell action. We construct asymptotically flat Euclidean supersymmetric non-extremal solutions with two independent rotations and an arbitrary number of electric charges, providing black hole saddles of the gravitational path integral that computes a supersymmetric index, and evaluate their action equivariantly. We find that these Euclidean saddles interpolate between supersymmetric extremal black holes and two-center horizonless microstate geometries. The interpolation involves dialing the temperature and implementing different analytic continuations. The corresponding on-shell action does not depend on temperature but is affected by the analytic continuations.
}

\newpage
\setcounter{page}{1}
\pagestyle{plain}

\tableofcontents

\vskip 1cm


\section{Introduction}

In the Euclidean approach to the gravitational path integral, saddle point contributions to the partition function are computed by evaluating the action on solutions to the classical equations of motion respecting the boundary conditions~\cite{Gibbons:1976ue}. One can then try to take short distance corrections into account.
 Whenever  the partition function can be computed by other methods, such as fundamental string theory or holography,  it is instructive to compare the results, so as to gain new insight into the relations between the macroscopic and microscopic descriptions of quantum gravity. While going beyond classical gravity is hard in general, in the presence of supersymmetry very precise comparisons become possible, including quantum corrections. A main goal in this subject is to determine exact partition functions computing quantum black hole entropies;  see for instance the reviews \cite{Sen:2007qy,Mandal:2010cj,Zaffaroni:2019dhb}.

In order to determine the different phases of the theory, one should in principle know all classical solutions satisfying the assigned boundary conditions and compare their Euclidean action in order to identify the dominant ones. Given the difficulty of constructing new explicit solutions, it is desirable to develop techniques allowing to obtain the on-shell action even when the  solution is not explicitly known, just using minimal information such as topological data and boundary conditions. In this way one could learn a priori what is the contribution to the gravitational partition function of the saddles under consideration. The work of Gibbons-Hawking \cite{Gibbons:1979xm}, giving the on-shell action of four-dimensional gravitational instantons in terms of their zero-dimensional `nuts' and two-dimensional `bolts', is a pioneering development in this direction.
Indications that for supersymmetric solutions this goal should be achievable come from holography: it is known that supersymmetric field theory partition functions in curved space have a very constrained dependence on the geometry and other background fields~\cite{Closset:2013vra}, hence the same should be true in the dual gravitational partition function, see e.g.~\cite{BenettiGenolini:2016tsn} for a discussion.

Recently, starting with~\cite{BenettiGenolini:2023kxp,Martelli:2023oqk}, it has been realized that the mathematical tool of equivariant localization plays a role in supergravity and is useful for computing on-shell actions (related earlier ideas can be found e.g.\ in~\cite{BenettiGenolini:2019jdz,Hosseini:2019iad}).
This technique, based on the Berline-Vergne-Atiyah-Bott (BVAB) formula in equivariant cohomology~\cite{BerlineVergne,Atiyah:1984px}, allows to reduce the integral over a manifold to the fixed-point locus of a symmetry. The integration space may be either the external spacetime manifold, or the compactification manifold of higher-dimensional supergravity (or a submanifold thereof). By now, equivariant localization has been applied to the computation of a variety of gravitational observables, including black hole entropies, holographic central charges, free energies and scaling dimensions of operators~\cite{BenettiGenolini:2023yfe,BenettiGenolini:2023ndb,Colombo:2023fhu,BenettiGenolini:2024kyy,Suh:2024asy,Couzens:2024vbn,Hristov:2024cgj,BenettiGenolini:2024xeo}.
 Investigations so far have focused on even-dimensional spaces, where Killing vectors can have isolated fixed points.

In this paper, we discuss equivariant localization of the supergravity action in odd-dimensional spaces for the first time, focusing on ungauged five-dimensional supergravity. Then we apply our results to evaluate the on-shell action of Euclidean solutions related to black holes and topological solitons. In odd dimension, the fixed locus of a Killing vector has at least dimension one (since it always has even codimension).  Therefore, in five dimensions it  makes sense to adopt the terminology of `nuts' and `bolts' introduced in~\cite{Gibbons:1979xm} for the four-dimensional case: we will use `nuts' for the fixed loci of minimal dimension one, and `bolts' for those of dimension three.

We provide a localization formula for the on-shell action of ungauged five-dimensional supergravity, depending on nut and bolt contributions, as well as on a set of boundary terms. In particular, the nut contributions are given by the holonomy of certain one-form potentials $\nu$ constructed from the gauge fields in the theory.

As first application we discuss a non-supersymmetric example: we consider the Euclidean continuation of the general charged and doubly-rotating black hole solution. The ${\rm U}(1)^3$ symmetry of the solution and the $S^3$ topology of the horizon lead us to identify {\it two} Killing vectors, each with {\it one} nut. Each of these vectors is a combination of two ${\rm U}(1)$ isometries, with the third ${\rm U}(1)$ generating the fixed circle at the nut. We show that the corresponding localization formula reproduces the correct expression for the on-shell action. 
 In order to reach the final expression, we need some information about the solution close to the nut and at infinity, analogously to the four-dimensional analysis of~\cite{Gibbons:1979xm}. A crucial feature of our odd-dimensional setup is that the one-form potential $\nu$ which is integrated over the nut should be everywhere regular. This condition is non-trivial since the nut is a circle that shrinks to zero size elsewhere in the space (we illustrate this specific feature with a simple example where the volume of $S^3$ is computed equivariantly).
 We also show that there is a specific scheme choice such that the boundary terms cancel out, and the on-shell action is entirely given by the nut contribution. 
 
The supersymmetric limit of the solution above gives smooth non-extremal solutions representing the black hole contribution to the Euclidean gravitational path integral computing a supersymmetric index, that we will denote supersymmetric black hole saddles. Recently, it has been understood how supersymmetric black holes contribute to the gravitational grand-canonical partition function with  supersymmetric boundary conditions, which takes the form of a refined Witten index~\cite{Witten:1982df}.  The main observation is that while in Lorentzian signature supersymmetric black holes are also extremal and therefore develop an infinitely long throat that makes a direct evaluation of the action ill-defined, in the Euclidean regime one can consider analytic continuations of the black hole that are supersymmetric but non-extremal. These contribute with a finite action that is independent of the inverse temperature $\beta$ and thus also gives the supersymmetric and extremal action upon sending $\beta\to \infty$.
While this was first clarified in the asymptotically AdS case \cite{Cabo-Bizet:2018ehj,Cassani:2019mms}, it has been realized that the asymptotically flat case~\cite{Iliesiu:2021are,Hristov:2022pmo,Boruch:2023gfn} allows for a somewhat simpler analysis, as one can choose an analytic continuation of the parameters such that the metric is real (while in the AdS case one finds complex metrics); this facilitates the regularity analysis.

Next we turn to a systematic analysis of supersymmetric solutions. Although we will take a general approach based on the classification of~\cite{Gauntlett:2002nw}, we will be mostly interested in black hole saddles. We are going to consider the Euclidean supersymmetric non-extremal counterparts of the Lorentzian solutions of~\cite{Breckenridge:1996is,Cvetic:1996xz,Chamseddine:1998yv}, and show how their on-shell action is evaluated through equivariant localization. We also show how the action is determined in larger classes of supersymmetric solutions.

In the presence of supersymmetry, the equivariant integral is fixed by fewer data than in the general case, similarly to what has been found in even dimension. However, we face a feature that makes the five-dimensional analysis different from the even-dimensional one: the `supersymmetric' Killing vector that is obtained as a bilinear of the Killing spinor is nowhere vanishing in  the Euclidean solutions of interest.
 This should be contrasted with the situation in even dimension, where the supersymmetric Killing vector can be used to localize the action to fixed-point contributions; for instance, in the supersymmetric non-extremal limit of the Kerr-Newman black hole this vector has a nut at the poles of the horizon two-sphere~\cite{Whitt:1984wk} (see e.g.~\cite{BenettiGenolini:2016tsn,BenettiGenolini:2019jdz,Bobev:2020pjk,Cassani:2021dwa,BenettiGenolini:2024xeo} for more general 4D solutions with a cosmological constant where the action is computed in terms of fixed points of the supersymmetric Killing vector). This leads us to consider a different vector for localizing: we assume the existence of an additional ${\rm U}(1)$ isometry that preserves the supercharge and use it to construct two Killing vectors with one nut.\footnote{For black hole saddles, these vectors are just the supersymmetric limit of the ones identified in the non-supersymmetric case mentioned above.} The supersymmetric solution is then characterized by harmonic functions on three-dimensional flat space~\cite{Gauntlett:2002nw}. We construct the one-form potential $\nu$ and show how the action integral is determined by the coefficients of the harmonic functions at their centers. We study in detail the case where there are just two centers and the whole solution has further enhanced ${\rm U}(1)^3$ symmetry. The final formula for the action gives a sum over the centers. We also obtain a characterization of the scheme choice ensuring cancellation of boundary terms that only relies on the supersymmetry algebra and on boundary conditions.
 
The setup above gives us the opportunity to discuss an alternative approach to the evaluation of the five-dimensional action: at least when the additional ${\rm U}(1)$ isometry is freely acting, we can reduce to four dimensions and then apply equivariant localization there. In this way, we are able to match our results with those obtained by localizing with respect to the supersymmetric Killing vector in four dimensions. We explicitly show this in the example of  the Kerr-Newman solution and its uplift to five dimensions, which is also captured by harmonic functions with two centers. We find that the sum over the two centers in the five-dimensional formula is mapped into a sum over the nuts in four dimensions.

Finally, we extend our analysis to five-dimensional supergravity coupled to an arbitrary number of vector multiplets. In this case, the supersymmetric non-extremal black hole saddles are not already available for arbitrary rotation and electric charges, so we construct them by directly solving the supersymmetry equations. Then we evaluate their action via equivariant localization and analyze their thermodynamic properties by taking suitable analytic continuations. 

We elaborate on an intriguing connection between the supersymmetric extremal black hole and horizonless topological solitons. Namely, we find that the supersymmetric black hole saddles with finite inverse temperature $\beta$ interpolate between the supersymmetric extremal black holes of~\cite{Breckenridge:1996is,Cvetic:1996xz,Chamseddine:1998yv},  which are obtained  taking $\beta \to \infty$ and continuing to Lorentzian signature,  and two-center horizonless solutions of~\cite{Giusto:2004id,Bena:2005va, Berglund:2005vb}, that we reproduce by taking the $\beta\to 0$ limit and turning to Lorentzian signature. 
 Although the supersymmetric on-shell action does not depend on $\beta$ and has fixed functional dependence on the remaining chemical potentials, the two limiting solutions require different analytic continuations of the parameters to make sense in Lorentzian signature. This implies that they arise for different values of the chemical potentials and carry a different action. The conserved charges also take different values.

The rest of the paper is organized as follows. In section~\ref{sec:equiv_5d} we illustrate how the on-shell action of five-dimensional supergravity can be computed via equivariant localization, present our non-supersymmetric example and discuss the scheme choice such that the boundary terms cancel out. In section~\ref{sec:susyaction}, we exploit the extra mileage of supersymmetry to evaluate the on-shell action equivariantly in a class of solutions with ${\rm U}(1)^3$ symmetry, including black hole saddles; we also compare with localization in four dimensions. In section~\ref{sec:multicharge} we extend our results to matter-coupled supergravity: we construct new supersymmetric non-extremal black hole saddles carrying multiple electric charges, compute their on-shell action via localization, discuss their thermodynamic properties, and show that they interpolate between the extremal black hole and the horizonless Lorentzian solutions. We conclude in section~\ref{sec:discussion}. Appendix~\ref{sec:4dexample} computes the action of the Kerr-Newman solution via equivariant localization,  as needed in section~\ref{sec:susyaction}. Appendix~\ref{sec:DiracMisner} contains a complementary analysis of the regularity of multi-charge saddles.


\section{Equivariant localization of the gravitational action}\label{sec:equiv_5d}

\subsection{BVAB formula in spaces with  boundary}

We start with a brief review of the Berline-Vergne-Atiyah-Bott (BVAB)  equivariant localization formula \cite{BerlineVergne,Atiyah:1984px} for a torus symmetry. A more extensive introduction can be found e.g.\ in~\cite{BerlineEtAlBook,Cremonesi:2013twh}. While the original BVAB formula is valid in compact spaces, here we will emphasize the modifications occurring for manifolds with a boundary, which will be needed in order to apply the formula to the evaluation of the gravitational action. 

 Let $(\mathcal{M},g)$ be a Riemannian manifold of dimension $D$, admitting a torus\footnote{Namely a product of U$(1)$'s, where the emphasis is on the fact that the group is compact.} isometry group. Let $\xi$ be a Killing vector field, generating an infinitesimal isometry of ${\cal M}$. We assume for the moment that $\mathcal{M}$ is compact with no boundary, we will add  boundary contributions at the end.  
We consider polyforms on $\mathcal{M}$, namely formal sums $\poly= \sum_n \poly_{(n)},$ where $\poly_{(n)}$ are forms of fixed degree $n=0,\ldots,D$. In particular, odd/even polyforms are made just of forms of odd/even degree.
The $\xi$-equivariant differential, 
\begin{equation}
\text d_\xi \,=\, \text d - \iota_\xi\,,
\end{equation}
sends odd/even polyforms into even/odd ones and squares to minus the Lie derivative along $\xi$, that is $\text d_\xi^2 = -\cal{L}_\xi$. 
Then, if we restrict to 
 polyforms satisfying  ${\cal L}_\xi \poly=0$,
the $\xi$-equivariant differential  can be used to define an equivariant cohomology.
A polyform $\poly$ is equivariantly closed if $\text d_\xi \poly=0$, while it
 is equivariantly exact if $\poly= \diff_\xi \Upsilon $ for some well-defined polyform $\Upsilon$. The equivariant cohomology  then consists of equivariantly closed modulo equivariantly exact polyforms.  This gives odd/even cohomology classes.
Note that the condition $\text d_\xi \poly=0$ leads to a set of recursive relations between forms of different degree (but same parity),  of the type
\be\label{descenteqs}
\iota_\xi \poly_{(n)} = \text d\poly_{(n-2)} \,.
\ee

We now consider the integral of a $\xi$-equivariantly closed polyform over a compact manifold $\mathcal{M}$ and sketch the localization argument showing that this only receives contributions from the fixed-point locus of the symmetry generated by $\xi$,  which we denote by ${\cal M}_0$.
First, the integral of a polyform over a $D$-dimensional manifold $\mathcal{M}$ picks by definition the component with the right degree to be integrated over,
$\int_{\cal M} \poly\, =\,  \int_{\cal M} \poly_{(D)}\,.
$
Hence by the ordinary Stokes' theorem  the integral of an equivariantly exact polyform over a compact manifold vanishes,
\begin{equation}
\int_{\cal M}\diff_\xi \Upsilon\, =\, \int_{\cal M} \text d\Upsilon_{(D-1)}\,= \,0\,,
\end{equation}
implying that integrals of equivariantly closed polyforms only depend on the equivariant cohomology class of the integrand. 

Next, we show that equivariantly closed polyforms on ${\cal M}$ are equivariantly exact on ${\cal M}\setminus {\cal M}_0$. Given the one-form dual to the vector $\xi$,  $\xi^{\flat} = \xi^\mu g_{\mu\nu}\text dx^\nu$, its equivariant differential  $\diff_\xi \xi^{\flat} = \diff \xi^{\flat} - |\xi|^2$ is invertible on ${\cal M}\setminus {\cal M}_0$, with inverse
\begin{equation}
\left( \diff_\xi \xi^{\flat}\right)^{-1} =-\frac{1}{|\xi|^2}\sum_{j=0}\left( \frac{\diff\xi^{\flat}}{|\xi|^2}\right)^j\,.
\end{equation}
It is easy to see that $\left(\diff_{\xi} \xi^{\flat}\right)^{-1}$ is equivariantly closed. It is convenient to consider the regulated space ${\cal M}_\epsilon$ obtained by removing infinitesimal tubular neighbourhoods about each connected component of ${\cal M}_0$, such that the original space is recovered in the $\epsilon\rightarrow 0$ limit. On the regulated space we can define the equivariant polyform
\begin{equation}
\Theta_{\xi} = \xi^{\flat} \wedge \left(\diff_{\xi} \xi^{\flat}\right)^{-1}\,,
\end{equation}
that by construction satisfies $\diff_\xi \Theta_{\xi} = 1$. An equivariantly closed polyform can now be written as
\begin{equation}\label{closed_is_exact}
\poly \,= \,\left(\text d_\xi \Theta_{\xi}\right)\poly \,=\, \text d_\xi \left( \Theta_{\xi} \poly \right)\,,
\end{equation}
showing that equivariantly closed polyforms on ${\cal M}$ are exact on ${\cal M}_\epsilon$. Then by the Stokes' theorem the integral of $\poly$ only receives contributions from $\partial {\cal M}_\epsilon$, that is the boundary of the infinitesimal neighbourhood about ${\cal M}_0$.  The BVAB  formula specifies what is such contribution in the $\epsilon\rightarrow 0$ limit: for any smooth $\poly$, it gives~\cite{BerlineVergne,Atiyah:1984px}\footnote{There is a simple refinement of the formula for the case where $\mathcal{M}$ has orbifold singularities, see e.g.~\cite{BenettiGenolini:2023kxp,Martelli:2023oqk,BenettiGenolini:2023ndb} for applications to supergravity. We do not consider it here.}
\begin{equation}
\label{eq:BVABformula}
\int_{\cal M} \poly \,=\, \int_{{\cal M}_0}\frac{\iota^*\poly}{ e_{\xi}({\cal N})}\,,
\end{equation}
where one should sum over the connected components of ${\cal M}_0$. On general grounds, each connected component must be of even codimension $2k$. The term $\iota^*\poly$ denotes the pullback over ${\cal M}_0$ of $\poly$ via the embedding $\iota: {\cal M}_0 \hookrightarrow {\cal M}$, while $e_{\xi}({\cal N})$ is the equivariant Euler form of the normal bundle ${\cal N} $ to ${\cal M}_0$ in ${\cal M}$. 
Quite generally, ${\cal N}$ splits into a direct sum of $k$ line bundles $L_i$, $i=1,\ldots,k$, which locally looks like $ \left(\mathbb R^2\right)^{k}$,
where each orthogonal $\mathbb R_i^2\subset \left(\mathbb R^2\right)^{k}$ has a corresponding angular coordinate $\phi_i$ with period $2\pi$, such that $\xi = \sum_{i=1}^{k} \epsilon_i\, \partial_{\phi_i}$. Then, the inverse of the Euler form has the expansion (see e.g.\ \cite{BenettiGenolini:2023ndb})
\be\label{exp_inv_Euler}
\frac{1}{e_\xi({\cal N})} \,=\, \frac{(2\pi)^k}{\Pi_{i=1}^k \epsilon_i } \left[ 1 - \sum_{1\leq i\leq k} \frac{2\pi}{\epsilon_i}\,c_1(L_i) + \sum_{1\leq i\leq j \leq k} \frac{(2\pi)^2}{\epsilon_i\epsilon_j} c_1(L_i)\wedge c_1(L_j)+ \ldots \right] \,,
\ee
 where $c_1(L_i)$ is the first Chern class of $L_i$.

Given the steps above, it is straightforward to adapt the BVAB formula to the case where the manifold ${\cal M}$ has a boundary, $\partial {\cal M}$ \cite{Couzens:2024vbn}. We assume that the latter does not contain fixed points of $\xi$.
The boundary $\partial {{\cal M}}_\epsilon$ introduced above is now given by the union of the boundaries of the infinitesimal neighbourhoods about the connected components of the fixed point locus, and the original boundary $\partial {\cal M}$. It follows that when integrating \eqref{closed_is_exact} and applying Stokes' theorem we need to include the boundary term on $\partial {\cal M}$, which reads
\be
 \int_{\partial {\cal M}} \Theta_{\xi} \poly \,=\, -\sum_{j=0}^{\left[\frac{D-2}{2}\right]} \int_{\partial {\cal M}} \eta \wedge \left(\diff \eta \right)^j\wedge \poly_{(D-2-2j)}\,,
\ee
where  we introduced the one-form
\be\label{eq:defeta}
\eta=\frac{\xi^{\flat}}{|\xi|^2}\,,
\ee
which is well defined on $\partial{{\cal M}}$ since by assumption $\xi$ does not vanish there.
Adding this boundary contribution to the BVAB formula \eqref{eq:BVABformula}, we arrive at the final expression
\begin{equation}\label{eq:BVAB_bdry}
\int_{\cal M} \poly \,=\, \int_{{\cal M}_0}\frac{\iota^*\poly}{ e_{\xi}({\cal N})}\,-\,\sum_{j=0}^{\left[{\frac{D-2}{2}}\right]} \int_{\partial {\cal M}} \eta \wedge \left(\diff \eta \right)^j\wedge \poly_{(D-2-2j)}\,.
\end{equation}

In the following we will be mostly interested in applying the equivariant localization formula to evaluate the gravitational action in odd-dimensional spaces. In particular, we will discuss in detail the five-dimensional case. In this case the connected components of the fixed locus of a Killing vector are either one-dimensional, or three-dimensional. By analogy with the four-dimensional terminology of Gibbons-Hawking~\cite{Gibbons:1979xm}, we will call `nuts' the former and `bolts' the latter.

\subsection{A simple example: the volume of $S^3$}\label{sec:volumeS3}

We now discuss a simple example (in compact space) that will be instructive in view of the applications to gravitational theories in the next sections. We consider a unit three-sphere $S^3$ and compute its volume, that is ${\rm Vol}_{S^3}=2\pi^2$, via equivariant localization. The metric is taken as
\be
\diff s^2 \,=\, \frac{1}{4}\diff \vartheta^2 + \sin^2\frac{\vartheta}{2}\, \diff \phi_1^2 + \cos^2\frac{\vartheta}{2}\, \diff \phi_2^2\,,
\ee
where $\vartheta\in [0,\pi]$, and $\phi_{1},\phi_2$  parameterize two circles $C_1$, $C_2$ of length $2\pi$. Here 
$S^3$ is described as a torus $C_1\times C_2$ foliated  over the interval parameterized by $\vartheta$: at the extremum  of the interval in  $\vartheta=0$ the circle $C_1$ shrinks to zero size, while $C_2$ remains finite. Conversely, at  $\vartheta=\pi$ it is $C_2$  that collapses while $C_1$ remains finite. 

We can localize with respect to the action of either one of the Killing vectors
\be
\xi_1 \,=\, \epsilon_1 \partial_{\phi_1}\,,\qquad\qquad \xi_2\, =\, \epsilon_2 \partial_{\phi_2}\,,
\ee with the result being independent of the choice made. Let us pick $\xi_1$: its fixed locus is the circle $C_2$ in $\vartheta=0$.
The following polyform is $\xi_1$-equivariantly closed and has the $S^3$ volume form as its top component:
\be
\Psi \,=\, \frac{1}{4} \sin\vartheta\, \diff\vartheta\wedge\diff \phi_1\wedge \diff\phi_2 \, + \,  \frac{\epsilon_1}{4}  \left(\cos\vartheta + \mu \right)\diff\phi_2\,,
\ee
where $\mu$ is an integration constant of the equivariant closure condition \eqref{descenteqs}; however this is not free, as regularity of the one-form  at $\vartheta=\pi$ (where $\xi_2$ vanishes) imposes $\mu=1$. The equivariant localization formula gives for the volume integral:
\be
{\rm Vol}_{S^3} \,=\, \int_{S^3} \Psi \,=\,  \frac{2\pi}{\epsilon_1}  \int_{C_{2,\vartheta=0}} \Psi_{(1)} \,=\,    \left(1 + \mu \right)\pi^2 \,,
\ee
which for $\mu=1$ is the correct result. 
In the computation we have kept $\mu$ generic so as to emphasize the importance of checking regularity everywhere on the manifold, not only close to the fixed locus of the vector chosen to localize. 

Another remark is that we could follow an alternate route, namely reduce to two dimensions and use localization there. In order to do so, we introduce new angular coordinates $\phi = \phi_1+\phi_2$, $\psi = \phi_2 - \phi_1$,
 so that the metric is expressed as
\be
\diff s^2 \,=\, \frac{1}{4}\left[ \diff \vartheta^2 + \sin^2\vartheta\, \diff\phi^2 + (\diff \psi + \cos\vartheta\, \diff \phi)^2 \right]\,,
\ee
while our Killing vector becomes
\be
\xi_1 \,=\, \epsilon_1 (\partial_\phi - \partial_\psi)\,.
\ee
Here $S^3$ is seen as the Hopf fibration $S^1\hookrightarrow S^3\to S^2$ of the circle of length $4\pi$ parameterized by $\psi$ over the unit two-sphere $S^2$ parameterized by $\vartheta,\phi$. The vector $\partial_\psi$ has constant norm and we can straightforwardly reduce along its orbit down to the two-sphere. The reduced volume integral is
\be\label{eq:volS3split}
{\rm Vol}_{S^3} \,=\,  \int_0^{4\pi} \!\diff\psi    \int_{S^2}  \frac{1}{8} \sin\vartheta \,\diff\vartheta\wedge \diff \phi 
\,,
\ee
while the vector reduces to
\be
\xi_1^{\rm 2D} \,=\,  \epsilon_1 \partial_\phi\,.
\ee
We can now apply equivariant localization in 2D using the equivariantly closed polyform
\be
\Psi^{\rm 2D} \,=\, \iota_{\partial_\psi}\Psi \,=\,  \frac18\sin\vartheta \,\diff\vartheta\wedge\diff \phi + \frac{\epsilon_1}{8} \,(\cos\vartheta + \mu)\,,
\ee
where now the integration constant $\mu$ is not constrained by any regularity condition as it appears in a zero-form.
On $S^2$, the vector $\xi^{\rm 2D}$ has isolated fixed points at both poles $\vartheta= 0$, $\vartheta = \pi$. So  \eqref{eq:volS3split} is given by a sum over two contributions, 
\be
{\rm Vol}_{S^3} \,=\,\int_0^{4\pi} \!\diff\psi   \int_{S^2} \Psi^{\rm 2D} \,=\,   4\pi \cdot\frac{2\pi}{\epsilon_1} \cdot \frac{\epsilon_1}{8} \left[   (1 + \mu)  -  (-1 + \mu) \right]  \,=\, 2\pi^2\,,    
\ee
with the integration constant trivially canceling out.

Comparing the two procedures above, we see that while the  two-dimensional integral is computed by a sum over both poles of the sphere, in the former three-dimensional computation one pole gives the fixed circle $C_2$ entering in the localization formula, while the other pole contributes indirectly, through the regularity condition for the pullback of the local one-form $\Psi_{(1)}$ on $C_2$.


\subsection{5D supergravity on-shell action as an equivariant integral}\label{sec:equiv5d}

We now apply the BVAB localization formula to the evaluation of the action of minimal five-dimensional supergravity on a (bosonic) solution. In Euclidean signature, the action reads

\begin{equation}\label{eq:actionminimal5d}
I_{\rm bulk} \,=\, - \frac{1}{16\pi }\int_{\cal M} \left( R\, \star_5 \!1 - \frac{1}{2}F\wedge \star_5 F + \frac{\ii}{3\sqrt{3}}A\wedge F \wedge F\right)\,,
\end{equation}
where $A$ is an abelian gauge field, and $F=\text dA$. It will also be useful to introduce the local three-form
\be\label{def_G_dualF}
G = \star_5 F - \frac{\ii}{\sqrt{3}}A\wedge F\,.
\ee
  The equations of motion read
\begin{equation}\label{eq:eqsofmotion5d}
\begin{aligned}
&R_{\mu\nu} - \frac{1}{2}F_{\mu\rho}F_\nu{}^\rho + \frac{1}{12}g_{\mu\nu} F_{\rho\sigma}F^{\rho\sigma}=0\,\,,\\
&\text dG=0\,.
\end{aligned}
\end{equation}

We consider a solution admitting a torus symmetry and we choose a vector field $\xi$ such that
\be
{\cal L}_\xi g \,=\, {\cal L}_\xi F \,=\, 0\,.
\ee
We will also pick a gauge such that 
\be
{\cal L}_\xi A \,=\, 0\,.
\ee

We now show that the on-shell action can be written as the integral of an equivariantly-closed polyform. Using the trace of the Einstein equation, \eqref{eq:actionminimal5d} can be written as
\begin{equation}\label{eq:action_after_einstein}
I_{\rm bulk} \,=\, \frac{1}{48\pi}\int_{\cal M} F\wedge G\,.
\end{equation}
The integrand $F\wedge G$ is the top-form $\Psi_{(5)}$ of an odd equivariantly closed polyform $\Psi = \Psi_{(5)}+ \Psi_{(3)} + \Psi_{(1)}$,  whose other components can be taken as
\begin{equation}\label{eq:polyform5d2}
\begin{aligned}
\Psi_{(3)} =&\, -\left(\iota_\xi A+\crho\right) G + \nu\wedge F \,,\\
\Psi_{(1)} =&\, -\left(\iota_\xi A+\crho\right)\nu + \iota_\xi \nu \,A\,,
\end{aligned}
\end{equation}
where the one-form $\nu$ is determined by the conditions 
\begin{equation}\label{eq:definitionnu5d_original}
\text d\nu = \iota_\xi G\,,\qquad\  \diff \iota_\xi \nu = 0\,,\qquad\ \crho\, \iota_\xi \nu = 0\,,
\end{equation}
and $\crho$ is a constant. The integrability condition $\diff \iota_\xi G =0$ for the first equation is satisfied using ${\cal L}_\xi G\equiv (\diff \iota_\xi + \iota_\xi \diff)\, G=0$ together with the Maxwell equation $\diff G=0$. Note that one-form potential $\nu$ is essentially the magnetic dual of the KK reduction of the gauge field $A$ along the orbits of $\xi$.

Whenever the vector $\xi$ vanishes somewhere, the contraction $\iota_\xi \nu$ -- which must be constant by the second in \eqref{eq:definitionnu5d_original} -- should be set to zero, so as to have a polyform $\poly$ that is regular along the shrinking direction. This  also solves the third in \eqref{eq:definitionnu5d_original}. Hence whenever $\xi$ has zeros, $\nu$ is defined by 
\be\label{eq:definitionnu5d}
\text d\nu = \iota_\xi G\,,\qquad\quad \iota_\xi \nu = 0\,.
\ee 
In this case the constant $\crho$ remains arbitrary. We will illustrate the role of $\crho$ in the example below, where it will be clear that it parameterizes different possible `schemes'.
We also stress that $\Psi$ should be everywhere regular, not just at the fixed locus of the vector that is used to localize, as made clear by the simple $S^3$ example of section~\ref{sec:volumeS3}.
 
Having constructed an equivariantly closed polyform whose top-component is the integrand of the on-shell bulk action, we can evaluate the latter by means of the BVAB localization formula \eqref{eq:BVAB_bdry}.
Expanding the formula, we find that the bulk integral can be expressed as
\be
\begin{aligned}\label{eq:localized_action_5d}
 I_{\rm bulk} \,&=\, \frac{1}{48\pi} \,\bigg[ \,\frac{(2\pi)^2}{ \epsilon_1 \epsilon_2 } \int_{{\rm nuts}} \iota^*\poly_{(1)}  +
\frac{2\pi}{\epsilon} \int_{{\rm bolts}}  \Big( \iota^*\poly_{(3)} - \frac{2\pi}{\epsilon}\, \iota^*\poly_{(1)}\wedge  c_1(L)  \Big) 
  \\[2mm]
&  \hspace{1.8cm}  -   \int_{\partial {\cal M}} \left(  \eta \wedge \poly_{(3)}+  \eta \wedge \diff \eta \wedge \poly_{(1)}  \right) \,\bigg] \,.
\end{aligned}
\ee
To this we must add the renormalized GHY boundary term,
\begin{equation}\label{eq:ghy_def}
I_{\rm GHY} \,=\, -\frac{1}{8\pi }\int_{\partial {\cal M}} \text d^4 x\left( \sqrt{h}\,{\cal K}- \sqrt{h_{\rm bkg}}\,{\cal K}_{\rm bkg}\right),
\end{equation}
where $h$ is the determinant of the induced metric and ${\cal K} $ is the extrinsic curvature of the boundary, while $h_{\rm bkg}$ and ${\cal K}_{\rm bkg}$ denote the corresponding quantities for a flat background with the same asymptotics as the solution under consideration, which cancels the divergences in the former term.
The full action then is 
\be
I = I_{\rm bulk} + I_{\rm GHY}\,.
\ee

Whenever the Killing vector has no fixed points and the regularity assumptions are satisfied, the action reduces to a boundary term on $\partial{\cal M}$. In this case one may have $\iota_\xi \nu\neq 0$, and the last in \eqref{eq:definitionnu5d} is solved by taking $\crho=0$. However, this would be essentially equivalent to using the Maxwell equation to reduce the action \eqref{eq:action_after_einstein} to the boundary term $I_{\rm bulk} =\frac{1}{48\pi}\int_{\partial\cal M} A\wedge \star_5 F \,,$ and then evaluating this term directly. 
 Therefore, here we will be interested in pursuing a different route: we will  choose a vector with zeros and find a choice of the constant $\crho$ such that the entire action is given by the contribution at the fixed locus.
In this case, since regularity at a nut requires $\iota_\xi A|_{\rm nut} = 0$, the nut contribution in \eqref{eq:localized_action_5d} can be expressed a little more explicitly as
 \be\label{eq:specify_Psi1}
 \int_{{\rm nuts}}\iota^*\poly_{(1)} \,=\, - \crho\, \int_{{\rm nuts}}\iota^* \nu\,,
 \ee
 so it is determined by the holonomies of the potential $\nu$ at the nuts.
 
 In the following, we illustrate the localization formula with a particularly relevant example which admits a supersymmetric limit. This will also be at the basis of the further developments in the next sections, where we will study more general supersymmetric solutions.
We will also provide the generalization to the case where the supergravity action includes an arbitrary number of vector multiplets, see section~\ref{sec:onshellaction_multicharge}.

\subsection{A non-supersymmetric example}\label{sec:nonsusy_ex}

We consider a general (non-supersymmetric and non-extremal) asymptotically flat black hole solution to the equations \eqref{eq:eqsofmotion5d}, which is included in the family first found in~\cite{Cvetic:1996xz}. We are going to compute the Euclidean on-shell action by means of the localization formula, showing in a concrete example how the localization argument works for odd-dimensional spaces, even in the absence of supersymmetry. 

The solution carries energy $E$, electric charge $Q$, and two independent angular momenta $J_1$, $J_2$. A convenient parametrization can be obtained by taking the ungauged limit  of the solution of minimal five-dimensional gauged supergravity given in~\cite{Chong:2005hr}. We also perform a Wick rotation to Euclidean time,  $t=-\ii \tau$, and focus on a real Euclidean section.  In coordinates $(r,\tau,\theta,\phi_1,\phi_2)$, the solution is given by\footnote{Besides switching off the gauging parameter, $g\to 0$, we have renamed some of the quantities appearing in~\cite{Chong:2005hr}  as  $\phi_{\rm there}= \phi_1$, $\psi_{\rm there} = \phi_2$, $\nu_{\rm there} = \ii\,\nu_1$ and $\omega_{\rm there} = \ii\,\nu_2$.  
 }
\begin{equation}\label{eq:cclpsolution}
\begin{aligned}
\text ds^2 \,&=\, \left(\text d\tau -\frac{2q\,\nu_1}{\rho^2}\right)\,\diff\tau - 2 q\,\frac{\nu_1\,\nu_2}{\rho^2}-\frac{2m\rho^2-q^2}{\rho^4}\left(\text d\tau + \nu_2\right)^2+\rho^2\left(\frac{\text dr^2}{\Delta_r}+ \text d\theta^2\right)\\[1mm]
\, &\quad \,+ \left(r^2-\aa^2\right)\sin^2\theta\,\text d\phi_1^2+\left(r^2-\bb^2\right)\cos^2\theta\, \text d\phi_2^2\,,
\\[1mm]
A\,&=\,  -\ii\, \frac{\sqrt{3}\,q}{\rho^2}\left( \text d\tau + \nu_2\right) + \ii \,\Phi\,\text d\tau\,,
\end{aligned}
\end{equation}
where 
\begin{equation}\label{eq:bh_functions}
\begin{aligned}
\nu_1\,&=\, \bb \sin^2\theta\, \text d\phi_1 + \aa\cos^2\theta \,\text d\phi_2\,\,,\quad\quad \nu_2 = \aa \sin^2\theta\, \text d\phi_1 + \bb \cos^2\theta\, \text d\phi_2\,\,,\\[1mm]
\Delta_r \,&=\, \frac{\left(r^2-\aa^2\right)\left(r^2-\bb^2\right)+q^2-2 \aa\bb q}{r^2}-2m\,,\\[1mm]
\rho^2\,&= \, r^2 -  \aa^2\cos^2\theta -  \bb^2 \sin^2\theta\,\,.
\end{aligned}
\end{equation}
The angular coordinates $\phi_{1},\phi_2$ are $2\pi$-periodic, while $\theta\in [0,\pi/2]$.
In order to obtain a real metric after Wick-rotating to Euclidean time, we have analytically continued the rotational parameters $a,b$ appearing in the original Lorentzian solution as $a= \ii \,\aa$ and $b= \ii\, \bb$. This is useful in order to study regularity of the Euclidean section, where we will assume $\aa,\bb$ real, as well as in order to satisfy the assumptions behind the equivariant localization argument. Note that the gauge field is then purely imaginary.\footnote{Although we will not do so here, one may find it convenient to also analytically continue the gauge field to a real field,
$A \to \ii A$, which would reabsorb the factor of $\ii$ appearing in front of the Chern-Simons term in \eqref{eq:actionminimal5d}. This would give a solution to the Euclidean five-dimensional supergravity discussed in~\cite{Sabra:2016abd}.}
The conserved charges are given in terms of the parameters $(m,q,\aa, \bb)$ by 
\begin{equation}\label{eq:cclpcharges}
E = \frac{3\pi}{4}m\,,\quad\quad Q = \frac{\sqrt{3}\pi}{4}q\,,\quad\quad J_1 = \frac{\pi \ii}{4}\left(2 \aa m + \bb q\right)\,,\quad\quad J_2 =\frac{\pi \ii}{4}\left(2\bb m + \aa q\right)\,,
\end{equation}
where $J_1$ and $J_2$ are associated with rotations generated by the Killing vectors $\partial_{\phi_1}$ and $\partial_{\phi_2}$, respectively.
For any conserved charge we can define a conjugate chemical potential, namely the inverse Hawking temperature $\beta$, the angular velocities $\Omega_{1}, \Omega_2$ and the electrostatic potential $\Phi$.
These can be computed in the standard way and read
\begin{equation}\label{eq:betacclp}
\beta = \frac{2\pi\,r_+\left[\left( r_+^2 -\aa^2 \right)\left( r_+^2 - \bb^2\right) - \aa\bb q\right]}{r_+^4 -\left( q- \aa\bb \right)^2}\,,
\end{equation}
\begin{equation}\label{ang_vel_nonsusy}
\Omega_1 = \ii \frac{\aa\left( r_+^2 - \bb^2\right)+ \bb q}{\left( r_+^2 - \aa^2\right)\left( r_+^2 - \bb^2\right) - \aa\bb q}\,\,,\quad\quad \Omega_2 = \ii \frac{\bb\left( r_+^2 - \aa^2\right)+ \aa q}{\left( r_+^2 - \aa^2\right)\left( r_+^2 - \bb^2\right) - \aa\bb q}\,\,,
\end{equation}
\begin{equation}\label{elec_pot_nonsusy}
\Phi = \frac{\sqrt{3}\, q\,r_+^2}{\left( r_+^2 -\aa^2 \right)\left( r_+^2 - \bb^2\right) - \aa\bb q}\,,
\end{equation}
where $r_+$ is determined by solving the equation $\Delta_r(r_+) =0$. In the Lorentzian solution, $r=r_+$ gives the event horizon, while in Euclidean signature it is the position where the solution caps off.\footnote{Note that the condition $\rho^2>0$, necessary for positive-definiteness of the metric, is non-trivial in the Euclidean solution:  we demand that this holds as long as $r> r_+$, namely we demand $r_+^2>  {\rm max}\,(\aa^2,\bb^2) $. In the supersymmetric case where $m=q$, this condition is  satisfied as long as $m>0$.
} Given the reality conditions chosen for the parameters, we see that in this Euclidean section $\beta$ and $\Phi$ are real while $\Omega_1,\Omega_2$ are purely imaginary.
 Finally,  the Bekenstein-Hawking entropy, computed as 1/4 the area of the three-sphere at $r=r_+$, is given by
\be\label{entropyE_minimal5d}
{\cal S} \,=\, \pi^2\,\frac{\left( r_+^2 -\aa^2 \right)\left( r_+^2 - \bb^2\right) - \aa\bb q}{2r_+}\,.
\ee

The angular velocities are such that the Killing vector 
\begin{equation}
W \,=\, \partial_\tau - \ii \Omega_1 \partial_{\phi_1} - \ii \Omega_2 \partial_{\phi_2}\,
\end{equation}
satisfies $W_\mu W^\mu |_{r_+} =0$ and thus vanishes at  $r=r_+$ (its analytic continuation $\ii W$ being the generator of the event horizon in Lorentzian signature). Regularity of the solution at $r_+$ requires  the Euclidean time to be compactified with period $\beta$, with the following twisted identification on the coordinates,
\begin{equation}
\left( \tau\,,\,\phi_1\,,\,\phi_2\right) \sim \left( \tau + \beta\,,\,\phi_1 - \ii \Omega_1 \beta\,,\,\phi_2 - \ii \Omega_2\beta\right)\,,
\end{equation}
in addition to the standard identifications
\begin{equation}
\left( \tau\,,\,\phi_1\,,\,\phi_2\right)\, \sim\, \left( \tau \,,\,\phi_1 +2\pi \,,\,\phi_2 \right)\, \sim\, \left( \tau \,,\,\phi_1  \,,\,\phi_2+2\pi \right)\,.
\end{equation}
Also,  in the expression for $A$  in~\eqref{eq:cclpsolution} we have fixed the gauge such that the field is regular at $r=r_+$, namely it satisfies
$
W^\mu A_\mu\big|_{r_+} =0\,.
$

With these identifications, the Euclidean solution we are considering topologically is $\mathbb R^2\times S^3$, with the first factor being parameterized by the  radial coordinate $r$ and  polar coordinate $\tau$, and the $S^3$ being described as already seen in section~\ref{sec:volumeS3}. 

 The isometry group is U$(1)^3$, generated by the Killing vectors $W$, generating rotations of period $\beta$ around the Euclidean thermal circle in $\mathbb R^2$, and $\partial_{\phi_{1}}$, $\partial_{\phi_{2}}$ in $S^3$.\footnote{One can introduce coordinates $\tilde\tau=\tau$, $\tilde\phi_1 = \phi_1+\ii \Omega_1\tau$, $\tilde\phi_2 = \phi_2 + \ii \Omega_2\tau$ which satisfy the untwisted identifications $$(\tilde\tau\,,\,\tilde\phi_1\,,\,\tilde\phi_2)\, \sim\, (\tilde\tau+\beta\,,\,\tilde\phi_1\,,\,\tilde\phi_2)\, \sim\, (\tilde\tau\,,\,\tilde\phi_1+2\pi\,,\,\tilde\phi_2)\, \sim\, (\tilde\tau\,,\,\tilde\phi_1\,,\,\tilde\phi_2+2\pi)\,.$$ In these coordinates, the vectors generating the U$(1)^3$ symmetry read  more simply as $W = \partial_{\tilde\tau}$, $\partial_{\tilde\phi_1}$, $\partial_{\tilde\phi_2}$.} 
As we have seen, each of these vectors vanishes on a three-dimensional hypersurface: at $r=r_+$, at $\theta =0$, and at $\theta = \pi/2$, respectively, hence these loci are bolts of the Euclidean geometry. 

We instead choose either one of the following Killing vectors for localizing the action:
\begin{equation}
\begin{aligned}
\label{eq:definexicclp}
\xi_N \,&=\, W +\ii\, (\Omega_1+\Omega_2)\, \partial_{\phi_1}\\[1mm]
\,&=\, \partial_\tau + \ii \,\Omega_2\, (\partial_{\phi_1}-\partial_{\phi_2})\,,
\end{aligned}
\end{equation}
which vanishes at the one-dimensional `north pole'  $\mn:  \{r=r_+\,,\ \theta = 0\}$, hence according to our terminology it has a nut at  $\mn$, or
\begin{equation}
\begin{aligned}
\xi_S \,&=\, W +\ii \left(\Omega_1+\Omega_2\right)\partial_{\phi_2} \\[1mm]
\,&=\, \partial_\tau - \ii \Omega_1 (\partial_{\phi_1}-\partial_{\phi_2})\,,
\end{aligned}
\end{equation}
which has a nut at  the `south pole' $\ms : \{ r=r_+\,\ \theta = \pi/2\}$. The result for the action integral must be independent of whether we choose to run the localization argument using $\xi_N$ or $\xi_S$. We are going to discuss in detail the case where we pick $\xi_N$, then we will comment on using $\xi_S$.
From the first line in~\eqref{eq:definexicclp}, we read that the equivariant parameters specifying the weights of the action of $\xi_N$ at the nut $\mn$ are 
\begin{equation}\label{eq:equivparamscclp}
\epsilon_1 = \frac{2\pi}{\beta}\,\,,\quad\quad \epsilon_2 = \ii \left(\Omega_1+\Omega_2\right)\,. 
\end{equation}

The localization argument states that the on-shell action reduces to a one-dimensional integral over $\mn$, up to a set of boundary terms. The first step in order to apply the localization formula \eqref{eq:localized_action_5d} is to compute the one-form potential $\nu$ by solving \eqref{eq:definitionnu5d}. We find it convenient to split the vector $\xi_N$ into the sum
\begin{equation}
\xi_N = V -   2\ii\Omega_2 \,U\,\,,\qquad\text{with}\qquad V =  \partial_\tau \,,\quad\quad U =  -\frac{1}{2} \left(\partial_{\phi_1} - \partial_{\phi_2}\right)\,,
\end{equation}
and consequently to split the one-form potential $\nu_N$ associated to $\xi_N$ as  $\nu_N = \nu_V -  2\ii\Omega_2\, \nu_U$, with 
\begin{equation}\label{eq:definenuVU}
\text d\nu_V = \iota_V G\,\,,\quad\qquad \text d\nu_U = \iota_U G\,,\qquad\quad \iota_{\xi_N} \nu_N = 0\,.
\end{equation}
This will allow us to easily compare with localization with respect to the other vector, $\xi_S$, which also is a combination of   $V$ and $U$.
We find the expressions
\begin{equation}
\label{eq:nucclp}
\begin{aligned}
\nu_V = &-\,\frac{\sqrt{3}\,\ii\,q}{\rho^2}\left[ \Omega_1 \Omega_2\,\rho^2 \,\text d\tau -\ii \Omega_2  \left(r_+^2 - \bb^2\right) \sin^2\theta\,\text d\phi_1 -\ii\Omega_1  \left( r_+^2 - \aa^2\right)\cos^2\theta\,\text d\phi_2  \right]\,,
\\[2mm]
\nu_U =&\,\ \frac{\sqrt{3}\, \ii \,q}{2\rho^2}\,  \bigg[\left(\ii(\Omega_1- \Omega_2)\,\rho^2 + \aa\cos^2\theta - \bb \sin^2 \theta + \frac{\Phi}{\sqrt{3}}\left(\bb\cos^2\theta - \aa \sin^2 \theta\right)\right)\text d\tau \\
&\qquad\qquad + \left(r^2-\bb^2\right) \sin^2\theta\, \text d\phi_1- \left( r^2 - \aa^2\right)\cos^2\theta\, \text d\phi_{2} \bigg]\,.
\end{aligned}
\end{equation} 
We emphasize that these are everywhere regular; in particular, they are regular at the nut $\mn$ as they satisfy $\iota_{\xi_N} \nu_V |_{\mn}=\iota_{\xi_N} \nu_U |_{\mn}=0\,$. 

We can now construct the three- and one-form in \eqref{eq:polyform5d2}, as well as the one-form $\eta$ in \eqref{eq:defeta}, and compute the different terms appearing in the localization formula \eqref{eq:localized_action_5d}. In order to compute the boundary terms, we will need the asymptotic behaviours for $r\to \infty$,
\be\label{eq:asympt_behav}
\eta \ \to\ \frac{1}{\ii\Omega_2} \left(\sin^2\theta\,\diff\phi_1 - \cos^2\theta\,\diff\phi_2 \right)\,,\qquad\qquad
\iota_{\xi_N} A \ \to\ \ii\, \Phi\,.
\ee
We find that the boundary term involving $\Psi_{(3)}$ is suppressed asymptotically.
On the other hand, since we have fixed a regular gauge, at the nut we have $\iota_{\xi_N} A |_{\mn}=0$.
Then the formula for the on-shell action reduces to a sum of three contributions,
\be\label{eq:local_action_bh}
 I \,=\, \frac{-\crho}{48\pi}\,\frac{(2\pi)^2}{ \epsilon_1 \epsilon_2 } \int_{\mn}\!\iota^*\nu_N   \, \,  +\,\, \frac{\ii\, \Phi+\crho}{48\pi} \int_{\partial {\cal M}}   \eta \wedge \diff \eta \wedge \nu_N   \, \, +\, \,I_{\rm GHY} \,.
\ee
Evaluating the boundary integrals using \eqref{eq:asympt_behav} we find
\begin{equation}\label{eq:bdrytermcclp}
 \frac{\ii\, \Phi+\crho}{48\pi}\int_{\partial{\cal M}} \eta \wedge \text d\eta \wedge\nu_N \,=\,-  \frac{\beta}{3}\left(\Phi - \ii \crho\right)Q\,,
\end{equation} 
\begin{equation}\label{eq:ghycclp}
I_{\rm GHY}\, =\, \frac{\beta}{3}\,E\,.
\end{equation}
The contribution from the fixed locus at $\mn$ can again be divided into the one coming from $\nu_V$ and the one from $\nu_U$. 
By pulling back on the nut we find
\begin{equation}\label{nutcclp_split}
\int_{\mn}\iota^*\nu_V\,=\, -8\Omega_1Q\,, \qquad\qquad
-2\ii \Omega_2\int_{\mn}\iota^*\nu_U\,=\, -8\Omega_2Q\,.
\end{equation}
Recalling the equivariant parameters given in \eqref{eq:equivparamscclp}, the full nut contribution to the action reads
\begin{equation}\label{eq:nutcclp}
\frac{- \crho}{48\pi} \frac{(2\pi)^2}{\epsilon_1\epsilon_2}\int_{\mn}\! \iota^*\nu_N\,=\, -\frac{\ii \crho}{3}\,\beta\,Q \,.
\end{equation}

Summing everything up we find the result
\begin{equation}\label{eq:result_action}
I = \frac{\beta}{3}\left(E-\Phi Q\right)\,,
\end{equation}
which is independent of the arbitrary coefficient $\crho$ introduced in~\eqref{eq:polyform5d2}, confirming that the latter represents a scheme choice.
This expression can be recast in a more familiar form by recalling the Smarr relation satisfied by the quantities given in \eqref{eq:cclpcharges}--\eqref{entropyE_minimal5d}, which reads
\begin{equation}
\beta E - \frac{3}{2}\,\mathcal S - \frac{3}{2}\left(\beta\Omega_1J_1 + \beta\Omega_2J_2\right) - \beta \Phi\,Q \,=\,0\,.
\end{equation}
Using this relation,~\eqref{eq:result_action} can be expressed as
\begin{equation}\label{QSE_general}
I = - \mathcal S +\beta\left(E -\Omega_1J_1 - \Omega_2J_2-\Phi\,Q\right)\,,
\end{equation}
that is the usual quantum statistical relation  identifying the on-shell action as a grand-canonical thermodynamic potential \cite{Gibbons:1976ue}.

Besides recovering the on-shell action via the localization formula, that is without explicitly evaluating the bulk integral, our main point here is that since the coefficient $\crho$ is arbitrary, we can choose it in such a way that the contribution from the asymptotic boundary terms, that is the sum of  \eqref{eq:bdrytermcclp} and \eqref{eq:ghycclp}, vanishes. This is achieved by taking
\begin{equation}\label{eq:cvarrho}
\ii \crho =   \Phi - \frac{E}{Q}\,.
\end{equation}
 Then, the full on-shell action arises as a nut contribution from $\mn$. 
 
 The lesson learned in this example, that there is a `localization scheme' such that the boundary terms cancel and the action arises solely from the fixed point set of the Killing vector, appears to be a general fact. Generically, in order to fix the constant $\crho$ to the desired value one should know some data of the solution that do not correspond merely to boundary conditions, such as the mass-to-charge ratio $E/Q$ in our example. 
 However, in the supersymmetric case this ratio is fixed by the supersymmetry algebra, so for supersymmetric solutions the constant $\crho$ is actually fixed by the choice of the potential $\Phi$, that is by boundary conditions only. Below we discuss the supersymmetric limit of the present solution. In the next sections we will present a  more systematic supersymmetric analysis. 
 
Let us also comment on working equivariantly with respect to $\xi_S$ instead of $\xi_N$. In terms of the vectors $V$ and $U$, we have $\xi_S = V +2 \ii\Omega_1 U$. The associated one-form potential is $\nu_S = \nu_V +2\ii \Omega_1 \nu_U$, with $\nu_V$ and $\nu_U$ being still given by \eqref{eq:nucclp}, and one can check that $\iota_{\xi_S}\nu_S =0$ as required by the equivariant closure condition. Fixing $\crho$ as in \eqref{eq:cvarrho} so that the boundary terms cancel (this is independent of whether we are using $\xi_N$ or $\xi_S$), the entire action is given by the nut contribution. The result must be the same as the one obtained earlier working equivariantly with respect to $\xi_N$. Since the equivariant parameters are also the same, we deduce that the following identity must hold
\be
\int_{{\cal M}_N}\iota^* \nu_N \,=\,  \int_{{\cal M}_S}\iota^* \nu_S \,,
\ee
Expressing $\nu_N$ and $\nu_S$ in terms of $\nu_V$ and $\nu_U$, this can also be written as
\be
\int_{{\cal M}_N}\iota^* \nu_V -2 \ii \Omega_2 \int_{{\cal M}_N}\iota^* \nu_U \,=\,  \int_{{\cal M}_S}\iota^* \nu_V +2 \ii \Omega_1 \int_{{\cal M}_S}\iota^* \nu_U \,.
\ee
Actually, a direct computation shows that these integrals over ${\cal M}_N$ and ${\cal M}_S$ are pairwise equal:
\begin{equation}
\int_{\mn}\iota^*\nu_V\,=\, -8\Omega_1Q \,=\,  2 \ii \Omega_1 \int_{{\cal M}_S}\iota^* \nu_U\,, \quad\quad
-2\ii \Omega_2\int_{\mn}\iota^*\nu_U\,=\, -8\Omega_2Q =  \int_{{\cal M}_S}\iota^* \nu_V\,.
\end{equation}
This implies that using the Killing vector $V$ and the associated potential $\nu_V$ is enough for obtaining the full action integral provided we integrate $\nu_V$ both on $\cal{M}_N$ and $\cal{M}_S$.

\subsubsection{Supersymmetric limit}\label{sec:susylimit_BH}
  
 The metric and gauge field \eqref{eq:cclpsolution} allow for a solution to the Killing spinor equation if $m=q$. Imposing this condition implies that the conserved charges \eqref{eq:cclpcharges} satisfy the linear relation 
\begin{equation}\label{eq:relationsuperalgebra}
E=\sqrt{3}\,Q\,,
\end{equation} 
as dictated by the superalgebra $\{\mathcal{Q},\overline{\mathcal{Q}}\}\sim E-\sqrt{3} Q$. However, it does not imply that the solution becomes extremal, as the inverse Hawking temperature \eqref{eq:betacclp} remains finite~\cite{Cabo-Bizet:2018ehj}. While the Riemannian metric is  regular, this non-extremal configuration has no well-definite Lorentzian counterpart for generic values of the parameter.

It is convenient to introduce the supersymmetric chemical potentials\footnote{These are just the ungauged versions of those defined in~\cite{Cabo-Bizet:2018ehj}. As in that reference, there is actually a second branch of expressions for the chemical potentials associated with the supersymmetric solution (arising because the equation $\Delta_r(r_+)=0$ has two solutions, see \eqref{eq:solq} below). Starting from the expressions in~\eqref{eq:cclpsusychempot}, the expressions for the other branch are obtained by sending $r_+ \to -r_+$  and flipping the overall signs. Since these other expressions lead to similar results, we will not explicitly display them in this paper.}
\begin{equation}
\label{eq:cclpsusychempot}
\begin{aligned}
\varphi \,=&\,\, \beta\big(\, \Phi - \sqrt{3}\,\big) \,=\, -2\sqrt{3}\pi\,\frac{\left(r_+ - \aa\right)\left(r_+ -\bb\right)}{2r_+ - \aa - \bb} \,,\\
\omega_1\, =&\,\, \beta\Omega_1\,=\, 2\pi\ii\,\frac{r_+ -\bb}{2r_+ - \aa -\bb} \,\,,\quad\quad \omega_2 \,=\, \beta\Omega_2\,=\, 2\pi\ii\, \frac{r_+ -\aa}{2r_+ - \aa -\bb}\,,
\end{aligned}
\end{equation} 
which satisfy
\be\label{eq:susyconstraint}
 \omega_1+\omega_2 \,=\, 2\pi \ii\,.
 \ee
 The quantum statistical relation~\eqref{QSE_general} then reads
 \begin{equation}
I = - \mathcal S -\omega_1J_1 -\omega_2J_2-\varphi\,Q\,.
\end{equation}

We then turn to the localization argument. We will work in the scheme where the boundary terms cancel out, given by the choice~\eqref{eq:cvarrho}. In the supersymmetric case, this reads
\be\label{eq:susy_c}
\ii \crho \,=\,   \Phi - \sqrt{3}\,,
\ee
hence it is fixed by boundary conditions only, as anticipated.\footnote{In four-dimensional supergravity, this amounts to a gauge choice, and corresponds to the `supersymmetric gauge' taken in~\cite{BenettiGenolini:2019jdz}.} 
The vector $V$ is the supersymmetric Killing vector arising as a bilinear of the Killing spinor, while $U$ generates an additional symmetry that preserves the supercharge; this will be discussed in detail in section~\ref{sec:susyaction}.
 The on-shell action is given by the sum $I = \mathcal I_N + \mathcal I_S$, where 
\begin{equation}\label{eq:cclpsusygravblocks}
\mathcal I_N = -\frac{\ii}{24\sqrt{3}}\,\frac{\varphi^3}{\omega_2}\,\,,\qquad\qquad \mathcal I_S = -\frac{\ii}{24\sqrt{3}}\,\frac{\varphi^3}{\omega_1}\,,
\end{equation}
 are the supersymmetric versions of \eqref{nutcclp_split}, \eqref{eq:nutcclp}.
The on-shell action can then be expressed as
\begin{equation}\label{eq:cclpsusyaction}
I \,=\, \frac{\pi}{12\sqrt{3}}\frac{\varphi^3}{\omega_1\,\omega_2}\,.
\end{equation}
This is also the result of~\cite{Cabo-Bizet:2018ehj} in the ungauged ($g\to 0$) limit, here recovered via equivariant localization. 
 
 \subsubsection{Contribution to the supersymmetric index and physical solutions}\label{contribindex}
 
The relation \eqref{eq:susyconstraint}, along with the fact that the solution supports a finite-size Euclidean thermal circle, provides the correct boundary conditions for the gravitational path integral to compute a supersymmetric index counting microstates~\cite{Cabo-Bizet:2018ehj,Iliesiu:2021are}. Indeed the gravitational path integral is interpreted as the grand-canonical partition function $Z= {\rm Tr}\, \rme^{-\beta (E -\Omega_1J_1-\Omega_2 J_2-\Phi Q )}  \,=\,{\rm Tr}\, \rme^{-\beta \{\mathcal{Q}, \overline{\mathcal{Q}} \} +\omega_1J_1+\omega_2 J_2+ \varphi Q }$, and the angular momentum terms can be written in the form 
$\text e^{\omega_1J_1+\omega_2J_2}= \text e^{2\pi \ii J_1+\omega_2(J_2-J_1)}= \left(-1\right)^F \text e^{\omega_2(J_2-J_1)}$, where $F$ is the fermion number and $\text e^{2\pi \ii J_1}=(-1)^F$. Hence we arrive at
\be\label{Zindex}
Z(\omega_2,\varphi)  \,=\,{\rm Tr}\, (-1)^F  \rme^{-\beta \{\mathcal{Q}, \overline{\mathcal{Q}} \} +\omega_2 (J_2-J_1) + \varphi Q}\,,
\ee
where the supercharge $\mathcal{Q}$ is chosen to commute both with $(J_2-J_1)$ and $Q$. This shows that the partition function $Z$ takes the form of a Witten index refined by chemical potentials $\omega_2,\varphi$. In particular, once expressed in the form \eqref{Zindex},  $Z$ does not depend on $\beta$, and the same holds for the Euclidean on-shell action $I$ providing a saddle-point contribution to the partition function, $Z \sim \rme^{-I}$. We also note that a sum over microstates of the form \eqref{Zindex} can be defined for complex $\omega_2$, $\varphi$, where usually one demands ${\rm Re}\, \omega_2 <0$, ${\rm Re} \,\varphi <0$ for convergence reasons.

 Having evaluated the action on the Euclidean solutions as above, we can then implement suitable analytic continuations  of the parameters and ask what sections of the solution represent actual saddles of the partition function~\eqref{Zindex} for the associated values of the chemical potentials.\footnote{When discussing saddles of a gravitational partition function and their thermodynamic interpretation, indeed, one may depart from the real section of the solution, see e.g.~\cite{Brown:1990fk,Witten:2021nzp} for discussions a priori of supersymmetry.}
 As a first thing, we observe that for the Euclidean solutions with real $\aa,\bb,q$ and $r_+$ discussed above, both $\omega_1$ and $\omega_2$ are purely imaginary. In the interpretation where the partition function  \eqref{Zindex} is given by a sum over microstates, a generic purely imaginary value of the chemical potential $\omega_2$ may be problematic as the sum may not converge. For this reason, it is not clear to us if the Euclidean solutions with real $\aa,\bb,q$ and $r_+$ should be regarded as saddles of \eqref{Zindex}. Nevertheless we note that $\varphi$ and $I$ are real in this case, and one can arrange for $\varphi<0$, $I>0$ (this is obtained by taking $r_+>\aa$, $r_+>\bb$).
  
 We then consider supersymmetric solutions with parameters $m=q$, $a=\ii \aa$, $b = \ii \bb$, $r_+$, related by the condition $\Delta_r(r_+)=0$, which can be expressed as 
 \be\label{eq:solq}
 q = r_+^2 -ab \pm \sqrt{- r_+^2(a+b)^2 }\,.
 \ee
 We see from this formula that in general some of the parameters should be taken complex. Let us in particular discuss the cases directly related to physical solutions, namely the cases where the conserved charges \eqref{eq:cclpcharges} and the entropy~\eqref{entropyE_minimal5d} are real. Reality of the conserved quantities implies that the parameters $q$, $a$, $b$ should all be taken real. From \eqref{eq:solq} we see that there are two possibilities: either $b = -a$, or $r_+^2<0$. 
  
  Taking $b = -a$ (and Wick-rotating back to real time $t=-\ii\tau$) we recover the extremal black hole, where we should also demand $q>a^2$ so that $r_+ = r_- = \sqrt{q-a^2}$ is real. From \eqref{eq:betacclp}--\eqref{elec_pot_nonsusy} it follows that $\beta = \infty$ while $\Omega_1=\Omega_2 = \Phi - \sqrt 3 =0 $. 
As for the supersymmetric chemical potentials, $\varphi$ is real and $\omega_1 = -\overline\omega_2$; moreover both ${\rm Re}\, \omega_2$ and $\varphi$ can be taken negative, while $I$ is real and positive. Hence we regard this as a good saddle, indicating that the supersymmetric and extremal black hole contributes to the index \eqref{Zindex}. 
  
On the other hand, considering the  case where $r_+^2<0$ we observe that the entropy \eqref{entropyE_minimal5d} would be purely imaginary, so it must vanish for the solution to be physical; this means that we must take $\left( r_+^2 +a^2 \right)\left( r_+^2 + b^2\right) +ab q =0$. This condition, together with $\Delta_r(r_+)=0$, gives\footnote{The other possible solutions are such that either $r_+$ or $q$ vanish.}
\be
q\,=\, -(2a+b)(a+2b)\,,\qquad\qquad r_+^2\,=\,-(a+b)^2\,.
\ee
  It also follows that $\beta=0$ while $\Omega_1,\Omega_2,\Phi$ diverge. The corresponding Lorentzian solution turns out to be the supersymmetric topological soliton discussed in~\cite{Chong:2005hr}, in the ungauged limit $g\to 0$. In this solution, the orbits of the angular Killing vector $\partial_{\phi_1}$ shrink to zero size as $\tilde r \equiv r^2- r_+^2\to 0$ and the geometry caps off with no horizon. It turns out that in the present ungauged setup one cannot avoid a conical singularity at the tip of the local geometry with polar coordinates $(\tilde r,\phi_1)$ and still have a non-trivial asymptotically flat solution. Allowing for a conical singularity introduces the quantization condition
  \be
  \frac{a+2b}{3(a+b)} \, = \, n \,\in\, \mathbb{Z} \,,\qquad \text{with $|n|>1$\,.}
  \ee
Then one can check that the four-dimensional hypersurface at fixed time has an $\mathbb{R}^4/\mathbb{Z}_{|n|}$ orbifold singularity of the ALE type located at $(r^2 = r_+^2$, $\theta = 0)$ as well as an $\mathbb{R}^4/\mathbb{Z}_{|1-n|}$ singularity at $(r^2 = r_+^2$, $\theta = \pi/2)$. The singularity analysis is similar to the one for the solutions of~\cite{Giusto:2004kj}, of which the present solution is a five-dimensional avatar.
Relying on our formulae despite the singularity, we find that the chemical potentials $\omega_1, \omega_2$, $\varphi$ are all finite and purely imaginary, their expressions being
 \be
 \omega_1 = 2\pi \ii n\,,\qquad  \omega_2 = 2\pi \ii (1-n)\,,\qquad \varphi =  -2\sqrt{3}\pi \ii n (2a+b)\,.
 \ee
It follows that the on-shell action \eqref{eq:cclpsusyaction} is also purely imaginary. We also notice that the microscopic sum defined by \eqref{Zindex} is blind to $J_1-J_2=2J_-$, since this takes integer eigenvalues both for fermionic and bosonic states and $\omega_2$ is a multiple of $2\pi \ii$. We will have more to say about this horizonless solution and its generalizations in section~\ref{sec:physical_prop}.

  \section{On-shell action of supersymmetric solutions\label{sec:susyaction}}
  
 In this section we use equivariant localization to compute the on-shell action of supersymmetric solutions to minimal five-dimensional supergravity, building on the classification of~\cite{Gauntlett:2002nw}. These solutions have at least one Killing vector (arising as a bilinear of the Killing spinor) and we will focus on the case where the latter is timelike. However, as we will show, the isometry generated by this Killing vector has no fixed points in the class of solutions of interest for us. We remedy this by focusing on supersymmetric solutions with a Gibbons-Hawking base-space, which possess at least one additional isometry: by introducing new vectors as linear combinations of the two isometries, we can identify those with a non-empty fixed-point set.

The section is organized as follows. In section~\ref{sec:gibbonshawkingansatz}, we first review the Lorentzian timelike supersymmetric solutions with Gibbons-Hawking base-space, then we discuss the analytic continuation to Euclidean signature. Next, we specify the asymptotic boundary behaviour of the metric and gauge field for the classes of solutions we intend to study. In section~\ref{sec:onshellaction} we compute the supersymmetric on-shell action, focusing on two-center Gibbons-Hawking bases. First, we demonstrate that the geometries of interest have a pair of Killing vectors that localize on one-dimensional nuts. Then we discuss regularity conditions, as well as the scheme choice such that the total boundary contribution to the on-shell action is cancelled and the result arises entirely from the nut contribution. In section~\ref{sec:exampleaction5d} we recover the example discussed in section~\ref{sec:susylimit_BH} from our more general analysis, while in section~\ref{sec:exampleaction4d} we make contact with localization in four dimensions.

\subsection{Supersymmetric solutions with an extra U(1) symmetry}\label{sec:gibbonshawkingansatz}

Bosonic supersymmetric solutions of the field equations admit a super-covariantly constant spinor $\epsilon$, called Killing spinor, satisfying a differential equation which arises from setting to zero the supersymmetry variation of the gravitino.
The local form of the bosonic supersymmetric solution is determined by analysing the differential forms that can be constructed as bilinears of the Killing spinor~\cite{Gauntlett:2002nw}. 

In particular, one can construct a Killing vector field $\tilde V$ as a one-form spinor bilinear. Introducing local coordinates such that $\tilde V = \partial_t$, the Lorentzian metric locally takes the form
\begin{equation}
\text ds^2 = - f^2\left( \text dt +\omega\right)^2 + f^{-1}\diff\hat s^2\,\,,
\end{equation}
where $\diff\hat s^2$ is a four-dimensional hyper-K\"ahler metric, and $\omega $ is a local one-form on the hyper-K\"ahler base. We are interested in considering supersymmetric solutions in the case in which $\tilde V$ is a timelike Killing vector, with $f>0$. 

The gauge field, and the associated field strength, can be expressed as
\begin{equation}\label{eq:susysolminimal}
\begin{aligned}
A \,=&\,\, \sqrt{3}\left[-f\left( \text dt + \omega\right) + \hat A + \zeta\, \text dt \right]\,\,,\\
F \,=&\,\, \text dA \,=\, \sqrt{3}\left[-\text df \wedge \left(\text dt + \omega \right)- f\text d\omega + \text d\hat A \right]\,,
\end{aligned}
\end{equation}
where $\hat A$ is a gauge field on the hyper-K\"ahler base, and $\zeta$ is a constant. 

We will now assume that the four-dimensional hyper-K\"ahler manifold admits an additional U(1) isometry whose Killing vector field, denoted as $\partial_\psi$, preserves its triholomorphic structure. This case was also analyzed in~\cite{Gauntlett:2002nw}. One can show that then the hyper-K\"ahler metric takes the Gibbons-Hawking form
\begin{equation}\label{eq:susysolminimalghbase1}
\text d\hat s^2\,= \, H^{-1} \left( \text d\psi + \chi\right)^2 + H\,\delta_{ij}\,\text dx^i\,\text dx^j\,,
\end{equation}
where $\chi=\chi_i\,\text dx^i$ is a one-form on a three-dimensional flat base-space $\mathbb R^3$ with coordinates $x^i$, $i=1,2,3$, determined by the equation
\begin{equation}\label{eq:defchi}
\star_3 \text d\chi\, =\, \text dH\,,
\end{equation}
where $H$ is a harmonic function on $\mathbb R^3$. Assuming further that the Killing vector $\partial_\psi$ generates a symmetry for the full five-dimensional spacetime, one can decompose the one-form appearing in the solution as
\begin{equation}\label{eq:susysolminimalghbase2}
\begin{aligned}
\omega \,=&\,\, \omega_\psi \left( \text d\psi +\chi\right) + \breve\omega\,\,,\qquad\qquad \breve\omega \,=\, \breve\omega_i \,\text dx^i\,,\\[1mm]
\hat A \,=&\,\, \frac{K}{H}\left( \text d\psi + \chi\right) + \breve A\,\,,\qquad\qquad \breve A \,=\, \breve A_i \,\text dx^i\,.
\end{aligned}
\end{equation}
All the functions and one-forms introduced above are determined in terms of a set of four harmonic functions $(H,K,L,M)$ with sources on $\mathbb R^3$,
\begin{equation}\label{eq:susysolminimalghbase3}
\begin{aligned}
\omega_\psi \,=\,\frac{K^3}{H^2}+\frac{3}{2}\frac{K\,L}{H}+M \,\,,\qquad\qquad f^{-1} \,=\, \frac{K^2}{H}+L\,,\\[1mm]
\star_3 \text d\breve\omega\,=\, H\text dM - M \text dH +\frac{3}{2}\left(K\text dL - L\text dK \right)\,\,,\qquad\qquad \star_3\text d\breve A \,=\, -\text dK\,.
\end{aligned}
\end{equation}
Then the solution depends on $4(s+1)$ parameters, which are the arbitrary coefficients of the four harmonic functions, with $s$ being the number of sources -- usually called `centers' in this context -- on the three-dimensional flat base-space. 
Some of these parameters will be fixed by imposing asymptotic conditions on the solution, or by requiring that the geometry is regular in the interior.  

Having reviewed the Lorentzian supersymmetry conditions, we now turn to Euclidean supersymmetry.  The general approach consists of regarding the spinors and their charge conjugate as independent, and complexifying all bosonic fields (including the metric). However, here we will focus on a subclass of the possible solutions, where the metric is taken real while the gauge field is purely imaginary. As we are going to show, these configurations are related to the supersymmetric configurations in Lorentzian signature summarized above via a simple analytic continuation.\footnote{This applies to the local supersymmetry conditions. As we will see, global regularity in Euclidean signature will lead us to supersymmetric non-extremal solutions that do not have a \hbox{Lorentzian counterpart.}} Hence we do not need to perform the  analysis of supersymmetry conditions again.

We start by performing the Wick rotation $t=-\ii \tau$. The five-dimensional Euclidean metric and gauge field become
\begin{equation}
\label{eq:euclideansusysolution}
\begin{aligned}
\text ds^2 \,=&\,\, f^2\left( \text d\tau + \ii \omega\right)^2 + f^{-1}\left( H^{-1}\left( \text d\psi + \chi\right)^2 + H \,\delta_{ij}\, \text dx^i \text dx^j\right)\,,\\[1mm]
\ii A\, =&\,\, \sqrt{3}\left[-f\left( \text d\tau + \ii \omega\right) + \ii \hat A + \zeta\, \text d\tau \right] \,,
\end{aligned}
\end{equation}
where the various quantities are determined by \eqref{eq:susysolminimalghbase2} and \eqref{eq:susysolminimalghbase3}. The period of the compactified Euclidean time, that is $\beta$, is determined by the standard argument that the solution must be regular in the interior. For the metric to be real, the one-form $\ii\omega$ must be real. This is achieved by requiring that the harmonic functions $K$ and $M$ become purely imaginary, while the others remain real. As a consequence, $\ii \hat A$ in \eqref{eq:euclideansusysolution} is real too. 

\paragraph{Asymptotics.}  We now turn to the asymptotic behaviour of the solution. The asymptotics of the multi-centered Gibbons-Hawking base-space are controlled by the function $H$~\cite{Gibbons:1979xm,Bena:2007kg,Gibbons:1978tef,Gibbons:1987sp,Gibbons:2013tqa},
\begin{equation}
H\,=\, h_0 + \sum_{a=1}^s\frac{h_a}{r_a}\,,
\end{equation}
where $r_a^2\equiv \sum_i \left( x^i - \overline x^i_a\right)^2$, and $\overline x^i_a$ denotes the location of the $a$-th center. In order to study the asymptotic metric, we introduce a system of coordinates that is centered at the origin of $\mathbb R^3$,
\begin{equation}\label{eq:coordinatesystem1}
x^1 = r\sin\vartheta\cos\phi\,\,,\quad\quad x^2 = r\sin\vartheta\sin\phi\,\,,\quad\quad x^3 = r\cos\vartheta\,,
\end{equation}
with $\vartheta \in [0,\pi]$ and $\phi \sim \phi + 2\pi$. We will consider solutions whose asymptotic metric and vector field at leading order for $r\to\infty$ are given by
\begin{equation}
\label{eq:asymptoticspacetime}
\begin{aligned}
\diff s^2 \rightarrow&\, \diff\tau^2 + \left( h_0 + \frac{h_+}{r}\right)^{-1}\! \left( \diff\psi + h_+\cos\vartheta\diff\phi\right)^2 + \left( h_0 + \frac{h_+}{r}\right)\!\left[ \diff r^2 + r^2\left( \diff\vartheta^2 + \sin^2\vartheta \diff\phi^2\right)\right]\!,\\[1mm]
\ii A \rightarrow&\, -\Phi \,\diff\tau\,,
\end{aligned}
\end{equation}
where $h_+ \equiv \sum_{j=1}^s h_j$. Then, the gauge parameter $\zeta$ appearing in \eqref{eq:euclideansusysolution} is fixed in terms of $\Phi$ as
\begin{equation}\label{eq:zeta}
\zeta = 1- \frac{\Phi}{\sqrt{3}}\,.
\end{equation}

If $h_0\neq 0$ the harmonic function $H$ goes like $H \rightarrow h_0$ at large $r$, and the four-dimensional hyper-K\"ahler base is asymptotic to the twisted product $S^1 \times \mathbb R^3$. In the literature these four-dimensional spacetimes are referred to as asymptotically locally flat (ALF). The ALF behaviour is obtained by requiring that the harmonic functions asymptotically tend to
\begin{equation}\label{eq:harmonicfunctionsalf}
H = h_0 + \mathcal O(r^{-1})\,\,,\quad\quad K = \mathcal O(r^{-2})\,\,,\quad\quad M = \mathcal O(r^{-2}) \,\,,\quad\quad L = 1+ \mathcal O(r^{-1})\,.
\end{equation}

On the other hand, if $h_0=0$ then  $H \rightarrow \frac{h_+}{r}$. This asymptotic behaviour is obtained by requiring the following conditions on the harmonic functions:
\begin{equation}\label{eq:harmonicfunctionsale}
H = \mathcal O(r^{-1})\,\,,\quad\quad K = \mathcal O(r^{-2})\,\,,\quad\quad M = \mathcal O(r^{-1}) \,\,,\quad\quad L = 1+ \mathcal O(r^{-1})\,\,.
\end{equation}
The global asymptotic structure depends on the periodicity $\Delta_\psi$ of the $\psi$ coordinate and the value of $h_+$.
If we take $\Delta_\psi=4\pi$, then we must take $h_+ \in \mathbb Z$, and the hyper-K\"ahler metric is asymptotically locally Euclidean (ALE), describing $\mathbb R^4/ \mathbb Z_{|h_+|}$ asymptotically. In particular, it is asymptotically flat for $\Delta_\psi=4\pi$ and $h_+=1$. 

In the following, we will focus on two-center solutions, with harmonic functions given by\footnote{
Not all harmonic function coefficients are physically independent, as the supersymmetry equations are left invariant by certain shifts, see e.g.~\cite{Bena:2007kg}. In particular, the function $K= \sum_{a=1}^s\frac{k_a}{r_a}$ can be reparameterized by performing the shift $K\rightarrow K + \lambda \, H$. We use this freedom to impose the constraint $\sum_{a=1}^s k_a = 0$, in accordance with \eqref{eq:harmonicfunctionsalf} and \eqref{eq:harmonicfunctionsale}. 
}
\begin{equation}\label{eq:harmonicfunctiongeneral}
\begin{aligned}
H\,=&\,\, h_0 + \frac{h_N}{r_N}+ \frac{h_S}{r_S}\,\,,\quad\quad\ \ii K \,=\, \frac{\mathsf k}{r_N}-\frac{\mathsf k}{r_S}\,\,,\\[1mm]
\ii M\,=&\,\,  \frac{\mathsf m_N}{r_N}+ \frac{\mathsf m_S}{r_S}\,\,,\qquad\ \  \qquad  L\, =\, 1+ \frac{\ell_N}{r_N}+ \frac{\ell_S}{r_S}\,.
\end{aligned}
\end{equation}
In order for all the parameters appearing in \eqref{eq:harmonicfunctiongeneral} to be real, we have performed an analytic continuation of the coefficients of the purely imaginary harmonic functions $K$ and $M$ ($k = - \ii \mathsf k$, $m = -\ii \mathsf m$, where $k$ and $m$ are the coefficients in Lorentzian signature).
For ALE spacetimes we need to take $h_0=0$ in \eqref{eq:harmonicfunctiongeneral}. On the other hand, harmonic functions for ALF spacetime have $h_0\neq 0$, but $\mathsf m_S + \mathsf m_N =0$, in accordance with \eqref{eq:harmonicfunctionsalf}.

Two-center solutions admit an additional U(1) isometry, with Killing vector $\partial_\phi$, generating rotations about the axis that connects the centers. We can always choose a set of coordinates in $\mathbb R^3$ such that both centers lie on the $x^3$-axis, with positions given by $\overline x^i_{N,S}= (0,0,\pm \delta/2)$, where the parameter $\delta$ denotes the distance between them. In this case, the two centers coincide with the north and south pole, respectively, of the two-sphere  parametrized by the coordinates $(\vartheta,\phi)$ of \eqref{eq:coordinatesystem1}. For this reason we will refer to them as north and south pole and denote them as $\mathcal M_N$ and $\mathcal M_S$, respectively. 


\subsection{Boundary terms and fixed-point contribution}\label{sec:onshellaction}

In this section, we  compute the on-shell action equivariantly for the class of solutions described above. We will limit ourselves to two-center solutions, leaving for  future work generalizations to more complex scenarios, such as harmonic functions with more than two centers or solutions with different asymptotic behaviours.

\subsubsection{Regularity in the bulk and parameters of the solution}\label{sec:onshellactionpreliminaries}

In order to apply the localization formula we need a Killing vector with fixed points. One may a priori consider the supersymmetric Killing vector $V \equiv -\ii \tilde V=\partial_\tau$, whose squared norm is given by  $g_{\tau\tau} = f^2$. In fact, for supersymmetric and extremal black holes the horizon three-sphere is a bolt for this vector. The converse is also true: if $f|_{\mathcal M_a}=0$, with $a=N,S$, the solution develops and infinitely long AdS$_2$ throat near the center, $f^2$ has a double pole and the solution is extremal~\cite{Boruch:2023gfn}. 
However, in the present paper we are mostly interested in non-extremal solutions, with a thermal circle of finite length $\beta$; extremal solutions are then obtained by taking the limit $\beta\to\infty$. We then require that $f$ is a non-vanishing function on the base-space. Given the expression for $f$ in \eqref{eq:susysolminimalghbase3}, this non-extremality condition is satisfied by choosing the coefficients of the harmonic functions such that
\begin{equation}\label{eq:harmonicfunctionL}
\ell_a= \frac{\mathsf k^2}{h_a}\,\,,\quad\quad a = N,S\,\,.
\end{equation}
So the function $L$ is completely fixed in terms of the other.
Then the value of $f$ at the north center is
\be
f_N\,\equiv \, f\big|_{{\cal M}_N} \,=\,  \frac{\delta}{\delta\left( 1+ \frac{h_0\kk^2}{h_N^2}\right)+ \frac{\kk^2h_+^2}{h_N^2h_S}}\,,
\ee
while the value $f_S$ at the south center  is obtained from this expression by exchanging $h_N$ and $h_S$. The extremal limit is obtained by sending $\delta\to 0$ (it will be clear later that this implies $\beta\to\infty$), so that the two centers coalesce; we see that in this limit we recover the vanishing value $f_N=f_S=0$.
 
 We therefore consider a different vector for localizing. 
 Using the additional ${\rm U}(1)$ symmetry, we can combine the two commuting isometries $\partial_\tau$ and $\partial_\psi$ and consider Killing vectors of the type
\begin{equation}\label{eq:kvsusy}
\xi_a = V + \Omega_a\, U\,, \qquad\text{with}\qquad V=  \partial_\tau \,\,,\quad U = \partial_\psi\,,
\end{equation}
where $\Omega_a$ is a parameter, and $a$ labels the possible choices. This choice is very natural for solutions with a Gibbons-Hawking base-space, since it generates an isometry that preserves the hyper-K\"ahler complex structures. 
The norm is given by
\begin{equation}
|\xi_a|^2\,=\, f^2\left( 1+ \Omega_a\,\ii \omega_\psi\right)^2 + \frac{\Omega_a^2}{f\,H}\,.
\end{equation}
This vanishes at the centers ${\mathcal M_a}$, where the harmonic function $H$ diverges, if we choose
\begin{equation}\label{eq:vanishingnorm}
\Omega_a = -\frac{1}{\ii \omega_\psi}\bigg|_{\mathcal M_a}\,\,,\quad\quad a= N,S\,,
\end{equation}
provided $ \omega_\psi$ remains finite at the centers.  The conditions ensuring this fix $\mm_{a}$ in terms of the remaining parameters,
\begin{equation}
\mm_N=-\frac{\mathsf k^3}{2h_N^2}\,,\hspace{1cm}\mm_S=\frac{\mathsf k^3}{2h_S^2}\,.
\end{equation}
Thus, the general form of the harmonic functions $H, L, K , M$ for both ALF and ALE solutions is
\begin{equation}
\label{eq:regular_harmonic}
\begin{aligned}
H \,=&\,\, h_0+\frac{h_N}{r_N}+ \frac{h_S}{r_S}\,\,,\qquad\qquad\qquad\qquad \ii K \,=\, \mathsf k\left( \frac{1}{r_N}-\frac{1}{r_S}\right)\,,\\[1mm]
L\, =&\,\, 1+ \mathsf k^2\left( \frac{1}{h_N\,r_N}+\frac{1}{h_S\,r_S}\right)\,\,,\quad
\ii M\,=\, -\frac{\mathsf k^3}{2}\left(\frac{1}{h_N^2\,r_N}-\frac{1}{h_S^2\,r_S}\right)\,,
\end{aligned}
\end{equation}
where we have already set $\mm_{0}=0$, as required from \eqref{eq:harmonicfunctionsalf} and \eqref{eq:harmonicfunctionsale}. We recall that for ALE solutions $h_0 = 0$ and we further note that for ALF solutions $h_N=h_S$.

Therefore, the Euclidean geometries of interest are characterized by four key quantities, which will play a role in the localization argument. Firstly, the boundary conditions for the solution comprise the period of the compactified Euclidean time $\beta$ and the holonomy of the gauge field around the asymptotic thermal circle, $\beta\Phi$. Additionally, these solutions possess a pair of Killing vectors of the form \eqref{eq:kvsusy}, each with fixed points at one of the two centers. These vectors are determined by $\Omega_N$ and $\Omega_S$, which encode the relevant local data around the fixed-point sets.
Below, by studying the regularity conditions of the solution about the centers we are going to relate these four quantities to the  parameters of the solution
\begin{equation}
\left( \beta\,,\Phi\,,\,\Omega_N\,,\,\Omega_S\right)\ \longleftrightarrow\ \left( \delta\,,\,\mathsf k\,,\,h_N\,,\,h_S\right)\,.
\end{equation}

\paragraph{Regularity of the metric at the centers.}
In order to study the regularity of the metric around the centers of the harmonic functions it is convenient to introduce two sets of spherical coordinates on the three-dimensional base-space $\mathbb R^3$, originating from either one of the two centers:
\begin{equation}\label{eq:coordinatearoundcenters}
x^1 = r_a \sin\vartheta_a\,\cos\phi \,\,,\quad\quad x^2 = r_a\sin\vartheta_a \,\sin\phi\,\,,\quad\quad x^3 = \overline x^3_a + r_a \cos\vartheta_a\,.
\end{equation}
 We can now solve eqs.~\eqref{eq:defchi}, \eqref{eq:susysolminimalghbase3} and determine the one-forms $\chi$ and $\breve\omega$. Locally they can be expressed in terms of the coefficients of the harmonic functions as
\begin{equation}\begin{aligned}
\chi =&\, \left( h_N \cos\vartheta_N + h_S \cos\vartheta_S\right) \diff \phi\,,\\[1mm]
\breve\omega =&\, \left( \breve\omega_N \cos\vartheta_N + \breve\omega_S \cos\vartheta_S\right) \diff\phi + \breve\omega_{\rm reg}\,,
\end{aligned}
\end{equation}
with
\begin{equation}\label{eq:omegaterms}
\begin{aligned}
\ii \breve\omega_{N} \,= &\,\,-\frac{h_0\kk^3}{2h_N^2} -\frac{3}{2}\mathsf k -\frac{\kk^3 h^3_+}{2h_N^2 h_S^2\delta}\,,\hspace{1cm}\ii \breve\omega_{S} \,=\, \,\frac{h_0\kk^3}{2h_S^2} +\frac{3}{2}\mathsf k +\frac{\kk^3 h^3_+}{2h_N^2 h_S^2\delta}\,,\\[1mm]
\ii\breve\omega_{\rm reg} \,=&\,\, \frac{\kk^3 h^3_+}{2h_N^2 h_S^2\delta}\left(\cos\vartheta_N +1\right) \left( 1-\frac{r_N + \delta}{r_S}\right)\diff\phi\,,
\end{aligned}
\end{equation}
where $\breve\omega_{\rm reg}$ denotes the regular part of $\breve\omega$, that vanishes on the entire $x^3$-axis. 
We notice that for both ALE and ALF spacetimes one finds $\breve\omega_N = -\breve\omega_S$.
We rewrite the metric \eqref{eq:euclideansusysolution} isolating the coordinate $\psi$, obtaining
\begin{equation}
\diff s^2 =Y\left( \diff\psi + \chi + \frac{\ii \omega_\psi f^2}{Y}\left( \diff\tau + \ii \breve\omega\right)\right)^2 + \frac{f}{HY}\left( \diff\tau + \ii \breve\omega\right)^2 + \frac{H}{f}\delta_{ij} \,\diff x^i \diff x^j\,,
\end{equation}
where 
\begin{equation}
Y = f^2\,(\ii\omega_\psi)^2 + \frac{1}{f\,H}\,.
\end{equation}
Next, we expand the metric at leading order around the north center. To do so, we introduce a radial coordinate $\tilde r_N = 2 r^{1/2}_N$ and consider the limit $\tilde r_N\rightarrow 0$. We obtain
\begin{equation}\label{eq:metricaroundnut}
\begin{aligned}
\diff s^2 \,& \underset{\tilde r_N\to 0}{\longrightarrow} \ \frac{f_N^2}{\Omega_N^2}\Big( \diff \psi + \left( h_N \cos\vartheta_N + h_S\right) \diff\phi - \Omega_N \left( \diff\tau + \ii \breve\omega_N\left( \cos\vartheta_N -1\right) \diff\phi\right)\Big)^2 \\[1mm]
&\ +\frac{h_N}{f_N}\left[\diff \tilde r_N^2 + \frac{\tilde r_N^2}{4}\left(\frac{\Omega_N^2}{h_N^2}\left(\diff \tau + \ii \breve\omega_N\left( \cos\vartheta_N -1\right)\diff \phi \right)^2 + \left( \diff\vartheta_N^2 + \sin^2\vartheta_N \diff\phi^2\right) \right)\right]\,.
\end{aligned}
\end{equation}
Regularity of this metric imposes 
\begin{equation}
\label{eq:omegaN}
\Omega_{N}=\frac{h_N}{\ii {\breve \omega}_N}\, ,
\end{equation}
which further implies that the center is the smooth origin of an $\mathbb R^4$ factor, with $(\vartheta_N,\phi,\tau)$ describing a round $S^3$ shrinking along the $\tilde r_N$ radial direction,  provided $\beta$ satisfies 
\begin{equation}\label{eq:inversehawkingtemperaturesusy}
\beta = 4\pi  \ii \breve\omega_N\,,
\end{equation} 
and the coordinates satisfy suitable identifications. 
In order to show that the full five-dimensional metric around the center is smooth, we introduce the new coordinates
\begin{equation}\label{eq:regularnutcoordinate}
\tilde\tau =\frac{\Omega_N}{2h_N}\tau \,\,,\quad\quad \phi_N = \phi-\frac{\Omega_N}{2h_N}\tau\,\,,\quad\quad \psi_N = \psi + h_+ \phi - \Omega_N \tau\,,
\end{equation}
with
\begin{equation}
\tilde r_N = \sqrt{y_1^2 + y_2^2} \,\,,\quad\quad \vartheta_N = 2\,{\rm arctan}\left( \frac{y_1}{y_2}\right)\,,
\end{equation}
which give\footnote{We note that when writing \eqref{eq:spaceatnut} we are ignoring mixed components of the metric,  $g_{\psi_N \phi_N}$ and $g_{\psi_N {\tilde\tau}}$, which are of order ${\cal O}\left({\tilde r}_N^2\right)$.} 
\begin{equation}\label{eq:spaceatnut}
\diff s^2  \, \underset{\tilde r_N\to 0}{\longrightarrow} \  \frac{f_N^2}{\Omega_N^2}\,\diff\psi_N^2+ \frac{h_N}{f_N}\left( \diff y_1^2 + y_1^2\,\diff\phi_N^2+ \diff y_2^2 + y_2^2\,\diff\tilde\tau^2\right)\,,
\end{equation}
where the compact coordinates $\tilde\tau$ and $\phi_N$ are $2\pi$-periodic. On the other hand, the compact coordinate $\psi_N$ is identified with period $\Delta_\psi$, which can be determined by studying the global properties of the specific solution. 

Using the coordinates introduced in \eqref{eq:regularnutcoordinate}, one can write the Killing vector that vanishes at the north pole as
\begin{equation}\label{eq:kvnorth}
\xi_N = \epsilon_1^N \partial_{\tilde\tau} + \epsilon_2^N \partial_{\phi_N}\,,
\end{equation}
where $\tilde\tau$ and $\phi_N$ are standard polar coordinates of period $2\pi$ that rotate each orthogonal copy of $\mathbb R^2$, and $\epsilon_{1,2}^N$ are the weights of such rotations. Namely, $\epsilon_{1,2}^{N}$ is the set of equivariant parameters that specifies the action of the Killing vector $\xi_{N}$ at the nut. As is evident from \eqref{eq:spaceatnut}, the tangent space to the nut comprises an $\mathbb R^2\oplus \mathbb R^2$ factor, and each point in $\mathbb R^2\oplus \mathbb R^2$ is associated with an $S^1$ parameterized by $\psi_N$ that remains invariant under the action of the Killing vector \eqref{eq:kvnorth}. 
The equivariant parameters can be explicitly computed by starting with the Killing vector given in \eqref{eq:kvsusy}. After performing the change of coordinate \eqref{eq:regularnutcoordinate}, we express the vector as shown in \eqref{eq:kvnorth}. Then, they are given by
\begin{equation}\label{eq:equivparams}
\epsilon_1^N = \frac{2\pi}{\beta}= - \epsilon_2^N\,.
\end{equation}
A similar expansion around the south pole gives 
\begin{equation}\label{eq:omegaS}
\Omega_S = - \frac{h_S}{\ii {\breve \omega}_N}\,,
\end{equation}
with equivariant parameters
\begin{equation}\label{eq:equivparams2}
\epsilon_1^S = \frac{2\pi}{\beta}= \epsilon_2^S\,.
\end{equation}

\paragraph{Regularity of the one-forms.} We require that the gauge field $A$ be regular at the nuts. We thus need to impose the conditions:
\begin{equation}\label{eq:reggaugevector}
\iota_{\xi_a}A\big|_{\mathcal M_a} \,=\, 0 \quad \Leftrightarrow \quad   \Phi   - \sqrt{3}  \left[ 1 -  f\left( 1+ \Omega_a\,\ii\omega_\psi \right) +  \Omega_a\frac{\ii K}{H}\right]_{\mathcal M_a}\!\!=
 0\,,\quad\ a = N,S.
\end{equation}
These two separate conditions should be met for a given value of $\Phi$. Using \eqref{eq:vanishingnorm}, \eqref{eq:omegaN} and \eqref{eq:omegaS}, the equations \eqref{eq:reggaugevector} have a unique solution 
\begin{equation}\label{eq:gaugechoice}
 \Phi\,=\, \sqrt{3}\left(1+  \frac{4 \pi \kk}{\beta} \right) \,,
\end{equation}
which fixes $\kk$.

We can now construct the equivariantly closed form \eqref{eq:polyform5d2} needed for the localizing the action. We are free to work equivariantly either with respect to $\xi_N$ or $\xi_S$. Let us consider $\xi_N$. The only missing ingredient is the evaluation of the one-form potential $\nu$ according to its definition \eqref{eq:definitionnu5d}.\footnote{From now on we drop the `$N$' from $\nu_N$ so as to simplify the notation.} Note that $\nu$ is defined locally, i.e.\ up to an exact term. This should be chosen so that $\nu$ is globally well defined.
A convenient way to determine $\nu$ is to split it into the sum
\begin{equation}
\nu = \nu_V +\Omega_N\, \nu_{U}\,\,,\quad\quad \diff \nu_V= \iota_V G\,\,,\quad\quad \diff \nu_{U} = \iota_{U} G\,,
\end{equation}
and use the relations given by the supersymmetry equations for the general solution with Gibbons-Hawking base-space \eqref{eq:euclideansusysolution}. 
A lengthy, but straightforward, computation gives
\begin{equation}\label{eq:nuV}
\nu_V = \sqrt{3}\,\ii V^\flat - f\,A- \zeta \,A\,,
\end{equation}
where $V^\flat = g_{\mu\nu}V^\nu\diff x^\mu$, and
\begin{equation}\label{eq:nuU}
\nu_{U} = -\sqrt{3}\,\ii \left[\varepsilon + \ii \left(\frac{K}{H}-f\omega_\psi \right)\left(\ii \breve A + \zeta \diff\tau \right) + f\frac{\ii K}{H}\left( \diff \tau + \ii \breve \omega\right)+\mu\right]\,,
\end{equation}
with
\begin{equation}
\diff \varepsilon = \star_3 \diff L\,,
\end{equation}
and where $\zeta$ is given in terms of $\Phi$ by \eqref{eq:zeta}.
In \eqref{eq:nuU} we used the freedom to add an exact term $\mu\equiv \mu_\tau \diff \tau + \mu_\psi \diff \psi$, for some constant coefficients $\mu_\tau$ and $\mu_\psi$. These are determined by solving the conditions
\begin{equation}\label{eq:regularitynususy}
\iota_{\xi_N} \nu = 0\,\,,\quad\quad \iota_{\xi_S} \nu\big|_S = 0\,,
\end{equation} 
where the first follows from regularity together with equivariant closure as in \eqref{eq:definitionnu5d}, while the second is required by regularity.
Using \eqref{eq:zeta}, \eqref{eq:vanishingnorm} and \eqref{eq:gaugechoice}, these give 
\begin{equation}
\begin{aligned}
\iota_{\xi_N}\nu\,=\, \frac{\ii}{\sqrt{3}}\left(\big(\Phi - \sqrt{3}\,\big)^2 - 3\Omega_N \left( \mu_\tau + \Omega_N \mu_\psi\right)\right)=0\,\,,\\[1mm]
\iota_{\xi_S}\nu\big|_S\,=\, \frac{\ii}{\sqrt{3}}\,\frac{\Omega_N}{\Omega_S}\left( \big(\Phi- \sqrt{3}\,\big)^2 - 3\Omega_S (\mu_\tau + \Omega_S \mu_\psi )\right) =0\,,
\end{aligned}
\end{equation}
which are solved by
 \begin{equation}
\mu_\tau = \frac{\Omega_N + \Omega_S}{3\Omega_N\,\Omega_S}\big(\Phi-\sqrt{3}\,\big)^2\,\,,\quad\quad \mu_\psi = -\frac{\big(\Phi-\sqrt{3}\,\big)^2}{3\Omega_N\,\Omega_S}\,. 
\end{equation}

Let us observe that the expressions for $\nu_V$ \eqref{eq:nuV} and $\nu_U$ \eqref{eq:nuU} are valid for any five-dimensional supersymmetric solution with Gibbons-Hawking base-space of the type reviewed in section \ref{sec:gibbonshawkingansatz}, regardless of the number of centers, or boundary conditions. However, the regularity conditions must be analyzed on a case by case basis. Here, using \eqref{eq:regularitynususy}, we have determined the explicit value of $\mu$ valid for the class of two-center solutions under study.

In summary, the regularity analysis gives us the following map between the coefficients of the harmonic functions and the  thermodynamic quantities that characterize the solution, now seen as the independent variables:
\begin{equation}\label{eq:mapcoeffchempot}
\mathsf k = \frac{\beta(\Phi -\sqrt{3})}{4\sqrt{3}\pi}\,\,,\quad\quad h_N = \frac{\beta \Omega_N}{4\pi}\,\,,\quad\quad h_S = - \frac{\beta\Omega_S}{4\pi}\,.
\end{equation}
where $\beta$ is related to the parameters of the solution via the first in \eqref{eq:inversehawkingtemperaturesusy} and \eqref{eq:omegaterms}. These quantities also determine the one-form $\nu$ appearing in the localization argument as discussed above.


\subsubsection{Boundary terms}\label{sec:actionboundaryterms}

We are finally ready to evaluate the on-shell action. We first calculate the boundary terms, consisting of those from the localization argument in the second line of \eqref{eq:localized_action_5d} and the renormalized GHY boundary term~\eqref{eq:ghy_def}.  The computation in this section will confirm that the value of the constant $\crho$ that cancels the total boundary contribution in supersymmetric solutions is the one given in \eqref{eq:susy_c}, even for more general asymptotic behaviours than the asymptotically flat one discussed in section~\ref{sec:nonsusy_ex}. 

The boundary terms from the second line of \eqref{eq:localized_action_5d} are all finite, so we can directly evaluate them at $r\to \infty$. The asymptotic behaviour of the fields was given in \eqref{eq:asymptoticspacetime}. It follows that the one-form $\eta = \xi_N^\flat/|\xi_N|^2$ asymptotically for $r\to\infty$ goes as
\begin{equation}
\eta\, \rightarrow\, \frac{h_0\, \diff\tau + \Omega_N \left( \diff\psi + h_+ \cos\vartheta\diff\phi\right)}{h_0 + \Omega_N^2}\,.
\end{equation}
Asymptotically, the non-suppressed contribution of $\Psi_{(3)}$ is dominated by the term proportional to $\star_5 F$, which goes as
\begin{equation}
\star_5 F \rightarrow - \sqrt{3}\,\ii \left(\diff\psi + h_+ \cos\vartheta\diff\phi\right)\wedge \star_3 \diff f^{-1}  \,\to\,  \sqrt{3}\,\ii \frac{\kk^2 h_+}{h_Nh_S} \sin\vartheta\, \diff \vartheta\wedge \diff\phi\wedge\diff\psi \,,
\end{equation}
where in the second step we used \eqref{eq:harmonicfunctionL}.
 This gives
\begin{equation}\label{eq:bdrytermalf1}
\frac{1}{48\pi}\int_{\partial \mathcal M}\left(\iota_{\xi_N} A + \crho\right) \eta\wedge \star_5 F\, =\, \frac{\Upsilon}{48\sqrt{3}\pi}\frac{h_0}{h_0+\Omega_N^2}\frac{\left(\Phi-\sqrt{3}\,\right)^2}{\Omega_N \,\Omega_S}\left(-\Phi + \ii \crho\right)\,,
\end{equation}
where $\Upsilon \equiv \frac{(h_0+\Omega_N^2)^{2}}{\Omega_N^2} \int_{\partial \mathcal M} \diff\tau \wedge \eta \wedge \diff\eta= - h_+  \int_{\partial \mathcal M} \sin\vartheta\,\diff\tau \wedge \diff \vartheta\wedge \diff\phi \wedge \diff\psi$.
Therefore, this contribution vanishes for ALE spacetimes, which are recovered by taking $h_0= 0$.

The boundary term involving $\Psi_{(1)}$ evaluates to
\begin{equation}\label{eq:bdrytermalf2}
\frac{1}{48\pi}\int_{\partial \mathcal M}\left(\iota_{\xi_N}A + \crho\right)\eta \wedge \diff \eta \wedge \nu \,=\,  \frac{\Upsilon}{48\sqrt{3}\pi}\frac{\Omega_N^2}{h_0 + \Omega_N^2}\frac{\left(\Phi-\sqrt{3}\,\right)^2}{\Omega_N \,\Omega_S}\left( -\Phi + \ii \crho\right)\,.
\end{equation}
Note that the sum \eqref{eq:bdrytermalf1}+\eqref{eq:bdrytermalf2} does not depend on $h_0$.

The renormalized GHY boundary term \eqref{eq:ghy_def} should be evaluated on a hypersurface at fixed $r$, with the limit  $r\to \infty$ being taken at the end of the computation. It can be written as
\begin{equation}
I_{\rm GHY} \,=\, -\frac{1}{8\pi}\int_{\partial \mathcal M}\diff^4x\left(n^\mu\partial_\mu\sqrt{h} - n_{\rm bkg}^\mu\partial_\mu\sqrt{h_{\rm bkg}}\right)\,,
\end{equation}
where the square root of the determinant of the induced metric is  $\sqrt{h}= f^{-1/2}H^{1/2}r^2\sin\vartheta$, the unit outward-pointing normal vector reads $n = f^{1/2}H^{-1/2}\partial_r$, while $\sqrt{h_{\rm bkg}}$ and $n_{\rm bkg}$ are the corresponding quantities for the reference background spacetime. 
 The result is
\begin{equation}\label{eq:ghyale}
\begin{aligned}
I_{\rm GHY}\,=\, -\frac{1}{8\pi}\lim_{r\rightarrow  \infty}\int_{\partial \mathcal M}\diff ^4x\, \frac{r^2}{2}\partial_r f^{-1} \sin\vartheta+ \mathcal O(r^{-1})\,=\, \frac{\Upsilon}{48\pi}\frac{\left(\Phi-\sqrt{3}\right)^2}{\Omega_N\,\Omega_S}
\,,
\end{aligned}
\end{equation}
which is independent of $h_0$.

It is now clear that the total boundary contribution, which is the sum of \eqref{eq:bdrytermalf1}, \eqref{eq:bdrytermalf2} and \eqref{eq:ghyale}, vanishes if we set $\crho$ as in \eqref{eq:susy_c} for both ALE and ALF spacetimes.

\subsubsection{Fixed-point contribution}\label{sec:actionnut}

Since the final result for the action is independent of $\crho$, we can assume the choice in  \eqref{eq:susy_c}, which as we have just demonstrated makes the total boundary contribution vanish for the entire class of solutions considered in this section. Consequently, from the first line of \eqref{eq:localized_action_5d} and \eqref{eq:specify_Psi1}, we have that the full on-shell action is given by a bulk contribution localized at the one-dimensional nut,
\begin{equation}\label{eq:actionfromnut1}
I 
 \,=\, -\frac{\left(2\pi\right)^2}{48\pi}\, \crho \int_{\mathcal M_N} \frac{\iota^* \nu}{\epsilon_1^N\,\epsilon_2^N}  \,.
\end{equation}

 The pullback of $\nu_V$ \eqref{eq:nuV} and $\nu_{U}$ \eqref{eq:nuU} on the north pole, using \eqref{eq:vanishingnorm} and \eqref{eq:gaugechoice}, reads
\begin{equation}\label{eq:pullbacknususy}
\begin{aligned}
\int_{\mathcal M_N}\iota^*\nu_V \,=&\,\, \Delta_\psi \big(\Phi - \sqrt{3} \,\big)\frac{K}{H}\,\Big|_{\mathcal M_N}\,,
\\[1mm]
\Omega_N\int_{\mathcal M_N} \iota^*\nu_{U} \,=&\,\,- \Delta_\psi \big(\Phi - \sqrt{3}\,\big)\frac{K}{H}\,\Big|_{\mathcal M_S}\,.
\end{aligned}
\end{equation} 
Note that the pullback of $\nu_U$ is given by a quantity evaluated at the south pole ${\cal M}_S$ and arises entirely from the exact form $\mu$ that we introduced around \eqref{eq:nuU} to ensure regularity of $\nu$. This highlights, once again, the central role of regularity conditions in the localization argument in odd dimensions. These conditions connect local information from the south pole to the integral evaluated at the north pole. 

As a consequence of \eqref{eq:pullbacknususy}, eq.~\eqref{eq:actionfromnut1} naturally splits into two separate blocks \\ $I = \mathcal I_N + \mathcal I_S$, where
\begin{equation}\label{eq:gravblocks}
\begin{aligned}
\mathcal I_N \,\equiv&\,\, -\frac{\left(2\pi\right)^2}{48\pi}\crho \int_{\mathcal M_N} \frac{\iota^*\nu_V}{\epsilon_1^N\,\epsilon_2^N} \,=\, \frac{\pi\Delta_\psi}{12\sqrt{3}} \frac{\left(\Phi-\sqrt{3}\,\right)^3}{\epsilon_1^N\,\epsilon_2^N\,\Omega_N}\,,\\[1mm]
\mathcal I_S \,\equiv&\,\, -\frac{\left(2\pi\right)^2}{48\pi}\crho\, \Omega_N \int_{\mathcal M_N} \frac{\iota^*\nu_{U}}{\epsilon_1^N\,\epsilon_2^N}\,=\, -\frac{\pi\Delta_\psi}{12\sqrt{3}} \frac{\left(\Phi-\sqrt{3}\,\right)^3}{\epsilon_1^N\,\epsilon_2^N\,\Omega_S}\,,
\end{aligned}
\end{equation}
with each block depending on $\Omega_N$ or $\Omega_S$, respectively. 

Finally, using \eqref{eq:mapcoeffchempot}, together with the value \eqref{eq:equivparams} of the equivariant parameters, we obtain the result for the on-shell action. We can express the action in terms of the coefficients of the harmonic functions, as 
\begin{equation}\label{eq:onshellactionfinal}
I \,=\,\mathcal I_N + \mathcal I_S=\,-\pi \Delta_\psi\frac{h_N + h_S}{h_N \,h_S}\,\mathsf k^3\,.
\end{equation}

\paragraph{The action as a sum of gravitational blocks.}

Note that $\mathcal I_S$ depends on the ratio $K/H$ computed at the south pole. It is then natural to ask whether we could compute $\mathcal I_S$ from a contribution that entirely comes from the south pole. This is indeed the case, as it is straightforward to show that
\begin{equation}
\mathcal I_S = -\frac{\left(2\pi\right)^2}{48\pi}\crho\int_{\mathcal M_S} \frac{\iota^*\nu_V}{\epsilon_1^S\epsilon_2^S}\,.
\end{equation}
Then, the full on-shell action can be expressed as a sum of terms that depend solely on $\nu_V$, computed both at the north and south pole,
\begin{equation}
I \,=\, \mathcal I_N + \mathcal I_S \,=\,  -\frac{\left(2\pi\right)^2}{48\pi}\crho\left( \int_{\mathcal M_N} \frac{\iota^*\nu_V}{\epsilon_1^N\epsilon_2^N} + \int_{\mathcal M_S} \frac{\iota^*\nu_V}{\epsilon_1^S\epsilon_2^S}\right)\,,
\end{equation}
which reminds of a sum over gravitational blocks~\cite{Hosseini:2019iad}. Recall that $V$ is the Killing vector obtained as a bilinear of the Killing spinor of the supersymmetric solution. This decomposition of the five-dimensional action extends what found in section \ref{sec:nonsusy_ex} for the black hole saddles and aligns with previous applications of equivariant localization in different contexts~\cite{Martelli:2023oqk,BenettiGenolini:2023yfe,Colombo:2023fhu,BenettiGenolini:2024kyy}, where gravitational blocks have been related to fixed-point contributions.


\subsection{Recovering the supersymmetric black hole saddles}\label{sec:exampleaction5d}

As a first application of the general formalism developed above, we reconsider the supersymmetric black hole saddle of section \ref{sec:susylimit_BH}, reviewing how it fits in the class of supersymmetric solutions with Gibbons-Hawking base-space and recovering the action \eqref{eq:cclpsusyaction}.
This will provide us with a useful guide in order to construct our more general multi-charge black hole saddles in section~\ref{sec:multicharge}. 

Following~\cite{Hegde:2023jmp}, we show that the solution \eqref{eq:cclpsolution}, after imposing supersymmetry as done in section~\ref{sec:susylimit_BH}, can be expressed in the form of a supersymmetric solution with a Gibbons-Hawking base-space \eqref{eq:euclideansusysolution} with two centers. First, the map 
\begin{equation}
\phi = \phi_1 + \phi_2\,\,,\quad\quad \psi = \phi_2 - \phi_1\,\,,\quad\quad \vartheta = 2\theta
\end{equation}
 relates the coordinates $(\theta,\phi_1,\phi_2)$ appearing in \eqref{eq:cclpsolution} with those appearing in the Gibbons-Hawking ansatz, where $\vartheta,\phi$ belong to the three-dimensional flat base space $\mathbb R^3$ of \eqref{eq:euclideansusysolution}. Starting from the global identifications of $\phi_1$ and $\phi_2$, we deduce that the angular coordinates $(\phi\,,\psi)$ have twisted identifications
\begin{equation}
\left( \phi\,,\,\psi\right) \,\sim\, \left( \phi + 2\pi\,,\psi+ 2\pi\right)\, \sim\, \left(\phi\,,\, \psi + 4\pi\right)\,,
\end{equation}
which fixes the period of $\psi$ as $\Delta_\psi = 4\pi$. 
For completeness, we report the full map $(x^1,x^2,x^3)\leftrightarrow (r,\vartheta,\phi)$ between the Cartesian coordinate on $\mathbb R^3$ and the coordinates of the solution \eqref{eq:cclpsolution} in the supersymmetric limit, that is
\begin{equation}
\begin{aligned}
x^1 \,=&\,\, \frac{1}{4}\sqrt{\left(r^2-r_+^2\right)\left(r^2-\left(-r_+ + \aa + \bb\right)^2\right)}\,\sin\vartheta\cos\phi\,,\\
x^2\,=&\,\, \frac{1}{4}\sqrt{\left(r^2-r_+^2\right)\left(r^2-\left(-r_+ + \aa + \bb\right)^2\right)}\,\sin\vartheta\sin\phi\,,\\
x^3\,=&\,\, \frac{1}{8}\left(2r^2-r_+^2 -\left(-r_++\aa+\bb\right)^2 \right)\,\cos\vartheta\,.
\end{aligned}
\end{equation} 
We see that the `horizon' at $r=r_+$ is mapped into a segment belonging to the $x^3$-axis, whose extrema coincide with the north and south pole of the two-sphere parametrized by $(\vartheta,\phi)$. These are the loci where the Killing vectors $\xi_N$, $\xi_S$ used in the localization argument vanish, and the centers of the harmonic functions are located. As above, we denote the north pole as $\mathcal M_N :\{r=r_+\,,\vartheta=0\}$, and the south pole as $\mathcal M_S : \{r=r_+,\,\vartheta=\pi\}$. The distance between the two centers is found to be 
\be
\delta \,=\, \frac{1}{4}\left( \aa + \bb\right)\left(2r_+ -\aa -\bb\right)\,.
\ee 
The full set of harmonic functions that specify the solution is given by \eqref{eq:regular_harmonic}, upon setting $h_0=0$. The parameters of the solution are determined in terms of the supersymmetric chemical potentials \eqref{eq:cclpsusychempot} by
\begin{equation}
h_N = -\ii\,\frac{\omega_2}{2\pi}\,\,,\qquad h_S = -\ii \,\frac{\omega_1}{2\pi}\,\,,\qquad \mathsf k = \frac{\varphi}{4\sqrt{3}\,\pi}\,.
\end{equation}
Remarkably, the supersymmetric linear constraint \eqref{eq:susyconstraint}
 translates into the asymptotic flatness condition $h_N + h_S \equiv h_+ =1$, recall the discussion under eq.~\eqref{eq:harmonicfunctionsale}.\footnote{A straightforward generalization of the solution is to allow for a simple orbifold, by taking $h_N+h_S\equiv h_+ \in \mathbb{Z}$. This gives a metric that asymptotically is $S^1_\beta\times \mathbb{R}^4/\mathbb{Z}_{|h_+|}$. The constraint between the angular velocities then becomes $\omega_1+\omega_2 = 2\pi \ii |h_+|$. Both the action and the Bekenstein-Hawking entropy are divided by $h_+$.}

We now turn our attention to the localization argument. The solution satisfies all our working assumptions for the equivariant computation of the on-shell action. For example, the two Killing vectors \eqref{eq:kvsusy} with nuts at the north or south center are determined by
\begin{equation}
\Omega_N = -\frac{1}{\ii \omega_\psi}\Big|_{\mathcal M_N} = \frac{4\pi h_N}{\beta}= -2\ii\Omega_2\,\,,\quad\quad \Omega_S = -\frac{1}{\ii \omega_\psi}\Big|_{\mathcal M_S} = -\frac{4\pi h_S}{\beta}= 2\ii\Omega_1\,,
\end{equation}
The on-shell action is ultimately given by the sum $I = \mathcal I_N + \mathcal I_S$, where $\mathcal I_{N,S}$ are defined in \eqref{eq:gravblocks}. For the choice of harmonic functions made in this section, the expressions provided in \eqref{eq:gravblocks} immediately reproduce the results in \eqref{eq:cclpsusygravblocks}, allowing us to conclude that the expression for the on-shell action presented in \eqref{eq:onshellactionfinal}
correctly reproduces the one in~\eqref{eq:cclpsusyaction}.

\subsection{Correspondence with localization in 4D}\label{sec:exampleaction4d}

As a second application, we consider the uplift to five dimensions of  four-dimensional supersymmetric Euclidean black hole saddles~\cite{Whitt:1984wk,Yuille:1987vw,Boruch:2023gfn}, and compute their action equivariantly using the general results derived above in this section. On the one hand, this will provide us with a class of five-dimensional ALF solutions, which displays a different asymptotic behaviour from the example considered in section~\ref{sec:exampleaction5d}. On the other hand, it will give us the opportunity to illustrate how equivariant localization in five dimensions matches localization in four dimensions upon dimensional reduction.
 In fact, an alternative approach to localization in odd dimensions for solutions possessing a  ${\rm U}(1)$ symmetry with no fixed points would consist of dimensionally reducing to an even-dimensional spacetime, and then applying localization there. In the context of supergravity, the effectiveness of equivariant localization in even dimensions has been demonstrated in a number of recent works, starting with~\cite{BenettiGenolini:2023kxp,Martelli:2023oqk}.
 
\paragraph{5D $\to$ 4D reduction.}

We take the following reduction ansatz for the metric and gauge field of five-dimensional minimal supergravity \hbox{along a Killing direction~$\partial_\psi$:}
\begin{equation}\label{eq:ansatzdimred}
\begin{aligned}
\diff s^2\, &=\, \diff s^2_{(4D)} + \left( \diff\psi + \mathcal A\right)^2\,,\\
A\, &=\, A_{(4D)}\,.
\end{aligned}
\end{equation}
Note that this is not the most general ansatz we could consider: while the generic ${\rm U}(1)$ reduction of five-dimensional minimal supergravity yields four-dimensional $\mathcal N=2$ supergravity coupled to a vector multiplet, here we want to focus on a reduction to pure $\mathcal{N}=2$ four-dimensional supergravity, truncating away the vector multiplet. As far as the bosonic sector is concerned, the latter truncation consists of setting to zero two four-dimensional  scalars, the dilaton from the metric and an axion  from the gauge field, as well as imposing the following duality relation between the gauge fields appearing in~\eqref{eq:ansatzdimred}:\footnote{In general in this section we denote four-dimensional quantities with the subscript $(4D)$. However, since the four-dimensional gauge field turns out to be equal to the five-dimensional one, 
 we will drop the subscript in $A_{(4D)}$ from now on.}
\begin{equation}\label{eq:dualization}
 \mathcal F = \frac{\ii}{\sqrt{3}}\star_4 F\,\,,
\end{equation}
where
${\cal F}=\diff {\cal A}$, $F = \diff A$, and  $\star_4$ is the Hodge-star built using the four-dimensional Euclidean metric. Substituting \eqref{eq:ansatzdimred}, \eqref{eq:dualization} into the five-dimensional field equations, we are left with four-dimensional field equations which are derived from the (Euclidean) pure $\mathcal{N}=2$ supergravity action,
\begin{equation}
I_{(4D)} \,=\, -\frac{\Delta_\psi}{16\pi}\int\left( R_{(4D)} \,\star_41 - \frac{2}{3}\,F\wedge \star_4 F\right) + I_{\rm GHY}\,,
\end{equation}
where $\Delta_\psi$ is the length of the reduction circle.\footnote{This factor is usually reabsorbed in the definition of Newton's gravitational constant of the reduced theory. However, since we chose to work in conventions such that $G_5=1$, we have $G_4= 1/\Delta_\psi$.}
 Note that, due to the need of dualizing away the KK gauge field $\mathcal A$ through~\eqref{eq:dualization}, substituting the ansatz \eqref{eq:ansatzdimred} into the five-dimensional action \eqref{eq:actionminimal5d} does not directly yield the correct four-dimensional action $I_{(4D)}$ of the consistent truncation. In order to address this, one should add to the reduced action  obtained by substituting \eqref{eq:ansatzdimred} a term proportional to $\int F \wedge  \mathcal{F}$, implementing the dualization upon varying with respect to $\mathcal{F}$.  More precisely, the actions are related as
\begin{equation}
I - \frac{\ii \Delta_\psi}{16\sqrt{3}\pi}\int F\wedge  \mathcal F\,=\, I_{(4D)} \,.
\end{equation}
When comparing the Euclidean on-shell actions in five and four dimensions, as we are going to do, it is important to include this dualization term.
Using Einstein's equations and \eqref{eq:dualization}, we can express the relation between the on-shell actions as
\begin{equation}\label{eq:action5dto4d}
I \,=\, \frac{I_{(4D)}}{2}\,.
\end{equation}

\paragraph{4D supersymmetric solutions.} Following~\cite{Gauntlett:2002nw}, a five-dimensional supersymmetric solution of the form \eqref{eq:ansatzdimred} is obtained by considering a solution with a Gibbons-Hawking base-space such that
\begin{equation}\label{eq:LandMIWP}
L = H\,,\quad\quad M = -\frac{1}{2}K\,.
\end{equation} 
Then, the four-dimensional metric and gauge field can be expressed as
\begin{equation}\label{eq:Eucl4d1}
\begin{aligned}
\diff s^2_{(4D)} \,=&\,\, H^{-1}f\left( \diff\tau + \ii \breve\omega\right)^2 + Hf^{-1}\delta_{ij}\diff x^i \diff x^j\,,\\[1mm]
\ii A \,=\,&\, \sqrt{3}\left(-f\left(\diff\tau + \ii \breve\omega\right) + \ii\breve A + \zeta\, \diff\tau \right)\,,
\end{aligned}
\end{equation}
where, according to \eqref{eq:defchi}, \eqref{eq:susysolminimalghbase2} and \eqref{eq:susysolminimalghbase3}, one finds that
\be\label{eq:Eucl4d2}
Hf^{-1}\,=\, K^2 + H^2\,\,,\quad\quad
\star_3\, \diff \breve\omega \,=\, 2\left( K \diff H - H \diff K\right)\,\,,\quad\quad \star_3\, \diff\breve A \,=\, - \diff K\,.
\ee

In Lorentzian signature, the supersymmetric solutions just described precisely correspond to the class of timelike supersymmetric solutions of four-dimensional pure $\mathcal{N}=2$ supergravity~\cite{Gauntlett:2002nw}.
 As first shown in~\cite{TOD1983241}, such four-dimensional solutions can be expressed in the form of the Israel-Wilson-Perjes (IWP) solution~\cite{Perjes:1971gv,Israel:1972vx} and are controlled by one complex function. 
 In the same way, the Euclidean solutions \eqref{eq:Eucl4d1}, \eqref{eq:Eucl4d2} match the Euclidean version of the timelike supersymmetric four-dimensional solutions, first discussed in \cite{Whitt:1984wk,Yuille:1987vw}. These solutions are determined by two independent harmonic functions, $\mathcal V$ and $\widetilde{\mathcal V}$. While $\mathcal V$, $\widetilde{\mathcal V}$ can a priori be complex \cite{Whitt:1984wk}, the metric is real only if both are taken real.  The agreement with \eqref{eq:Eucl4d1}, \eqref{eq:Eucl4d2} follows from identifying
\be\label{eq:mapV_H_K}
\mathcal V \,=\, H - \ii K\,,\quad\quad \widetilde{\mathcal V}\,=\, H + \ii K\,.
\ee
Recall that our reality conditions are such that $H$ is real while $K$ is purely imaginary, hence $\mathcal{V}$ and $\widetilde{\mathcal{V}}$ are real and independent.
Then the four-dimensional metric and gauge field can be written as
\begin{equation}
\begin{aligned}
\label{eq:IWPmetric}
\diff s_{(4D)}^2\,&=\, \frac{1}{\mathcal V\widetilde{\mathcal V}}\left( \diff \tau + \ii\breve\omega\right)^2 + \mathcal V\widetilde{\mathcal V}\,\delta_{ij}\, \diff x^i \diff x^j\,,\\[1mm]
\ii A \,&=\, \sqrt{3}\,\bigg(-\frac{\mathcal V + \widetilde{\mathcal V}}{2\mathcal V\widetilde{\mathcal V}}\left( \diff \tau + \ii \breve\omega\right) +\ii \breve A   +\zeta\, \diff\tau\bigg)\,, 
\end{aligned}
\end{equation}
where the local one-forms $\breve\omega$, $\breve A$ are expressed in terms of $\mathcal V$ and $\widetilde{\mathcal V}$ as
\begin{equation}
\begin{aligned}
\star_3\, \diff \breve\omega= -\ii\,\big(\,\widetilde{\mathcal V}\,\diff \mathcal V-\mathcal V\, \diff \widetilde{\mathcal V}\,\big) \,\,,\qquad\qquad 
\star_3\,\diff\breve A = -\frac{\ii}{2}\,\diff\big(\,\mathcal V - \widetilde{\mathcal V}\,\big)  \,.
\end{aligned}
\end{equation}

\paragraph{The 4D Euclidean saddle and its action.} A smooth and non-extremal Euclidean black hole saddle of the four-dimensional supergravity theory is obtained by choosing the harmonic functions as~\cite{Whitt:1984wk,Yuille:1987vw,Boruch:2023gfn}
\begin{equation}\label{eq:choiceVtildeV}
\mathcal V = 1 + \frac{q}{r_N}\,\,,\qquad\qquad \widetilde{\mathcal V} = 1 + \frac{q}{r_S}\,.
\end{equation}
Then, smoothness of the metric and the gauge field requires that the period of the compactified Euclidean time $\beta$ and the electrostatic potential $\Phi$ read
\begin{equation}\label{eq:beta_Phi_4D}
\beta = \frac{4\pi q\left( q+\delta\right)}{\delta}\,,\qquad\qquad\Phi=\frac{\sqrt{3}\left(2q+\delta\right)}{2\left(q+ \delta\right)}\,.
\end{equation}

As  verified in~\cite{Whitt:1984wk,Boruch:2023gfn,Hegde:2023jmp}, there is a change of coordinates that maps the Euclidean IWP solution \eqref{eq:IWPmetric}--\eqref{eq:choiceVtildeV} to the supersymmetric non-extremal limit of the Euclidean Kerr-Newman solution.

The four-dimensional on-shell action is given by a sum of a nut and an anti-nut contribution~\cite{Whitt:1984wk}. In appendix~\ref{sec:4dexample}, we compute the action of the generic electrically-charged Euclidean Kerr-Newman solution via equivariant localization, and recover the action of the solution above by taking the supersymmetric limit. 
There, we use the Killing vector $V = \partial_\tau$, which has two isolated fixed points at the poles of the horizon's two-sphere. The four-dimensional on-shell action  can then be written as the sum
\begin{equation}\label{eq:action4dsum}
I_{(4D)} = \frac{\pi\Delta_\psi}{6}\left(\frac{\Psi_{(0)}\big|_+}{\epsilon_1^+\,\epsilon_2^+}\,+\,\frac{\Psi_{(0)}\big|_-}{\epsilon_1^-\,\epsilon_2^-} \right)\,,
\end{equation}
where $\Psi_{(0)}$ is the zero-form component of the equivariantly-closed polyform whose top-form is the four-dimensional on-shell action, and the $\pm$ symbols denote the north and south pole of the horizon's two-sphere (see appendix~\ref{sec:4dexample} for details). 

\paragraph{Comparison with 5D computation.} In order to compare equivariant localization in five and four dimensions, we now compute the same four-dimensional action $I_{(4D)}$ using the five-dimensional uplifted   solution. In section~\ref{sec:actionnut} we derived the expression for the five-dimensional on-shell action for two-center solutions as a sum of two terms denoted as $\mathcal I_N$ and $\mathcal I_S$, with their expression being given in \eqref{eq:gravblocks}. We are going to show that these map, through the relation \eqref{eq:action5dto4d}, into the four-dimensional fixed-point contributions in \eqref{eq:action4dsum} as 
\begin{equation}\label{eq:map5d4dblocks}
\mathcal I_N \,=\, \frac{\pi\Delta_\psi}{12}\,\frac{\Psi_{(0)}\big|_+}{\epsilon_1^+\,\epsilon_2^+}\,\,,\qquad\qquad \mathcal I_S \,=\, \frac{\pi\Delta_\psi}{12}\,\frac{\Psi_{(0)}\big|_-}{\epsilon_1^-\,\epsilon_2^-}\,.
\end{equation} 

The five-dimensional solution is of the form \eqref{eq:ansatzdimred}, \eqref{eq:LandMIWP}--\eqref{eq:Eucl4d2}. From~\eqref{eq:mapV_H_K}, \eqref{eq:choiceVtildeV}, we obtain the harmonic functions
\begin{equation}\label{eq:HandKIWP}
H =\, 1 + \frac{q}{2}\left(\frac{1}{r_N}+ \frac{1}{r_S}\right)\,,\qquad\qquad \ii K = -\frac{q}{2}\left( \frac{1}{r_N}-\frac{1}{r_S}\right)\,.
\end{equation}
While the four-dimensional metric is asymptotically $S^1_\beta\times \mathbb{R}^3$, the five-dimensional metric asymptotically for $r\to\infty$ reads
\begin{equation}
\diff s^2 \ \to\ \diff\tau^2 + \diff r^2 + r^2 \left( \diff\vartheta^2 + \sin^2\vartheta \diff\phi^2\right)+ \left( \diff\psi + h_+ \cos\vartheta \diff\phi\right)^2\,,
\end{equation}
where one can take the period of $\psi$ as $\Delta_\psi = 4\pi h_+$ so that the hypersurfaces at fixed $\tau$ and $r$ topologically are three-spheres. Thus, this is an example of solution with ALF Gibbons-Hawking base-space, which is different from the one considered in section~\ref{sec:exampleaction5d}.

The expressions \eqref{eq:beta_Phi_4D} for the temperature and the electrostatic potential are consistent with \eqref{eq:inversehawkingtemperaturesusy} and \eqref{eq:gaugechoice}, meaning that a smooth five-dimensional Euclidean saddle corresponds to a smooth four-dimensional one. 

Unlike what happens for the four-dimensional geometry, in the five-dimensional solution $V=\partial_\tau$ has no fixed points. However, we can combine it with the generator $\partial_\psi$ of the compactified direction to form a pair of Killing vectors with a one-dimensional nut, in accordance with the discussion of section~\ref{sec:onshellactionpreliminaries}. These two vectors were defined in \eqref{eq:kvsusy} and read
\begin{equation}
\xi_N= \partial_\tau + \frac{2\pi q}{\beta}\,\partial_\psi \,\,,\qquad\qquad \xi_S = \partial_\tau - \frac{2\pi q}{\beta}\,\partial_\psi \,.
\end{equation}

It is now straightforward to verify that \eqref{eq:map5d4dblocks} holds. Indeed, the equivariant parameters can be mapped as $\epsilon_{1,2}^N= \epsilon_{1,2}^+$ and $\epsilon_{1,2}^S= \epsilon_{1,2}^-$, while the parameter $c$ specifying the localization scheme takes the same expression both in five and four dimensions, $\ii c= \Phi - \sqrt{3}$. A less straightforward correspondence holds between the integral of the pullbacks of $\nu_V$ and $\nu_U$ on $\mathcal M_N$, computed in \eqref{eq:pullbacknususy}, and the value of its four-dimensional analog $\nu_{(4D)}$  (cf.\ \eqref{eq:iotaAandnu4d}) at the nut and anti-nut. One obtains
\begin{equation}
\int_{\mathcal M_N} \!\!\!\iota^*\nu_V = \Delta_\psi\,\nu_{(4D)}\Big|_+ = - 2\sqrt{3}\pi\ii \Delta_\psi\frac{q}{\beta}\,,\quad\quad\Omega_N\! \int_{\mathcal M_N} \!\!\!\iota^*\nu_{U} = -\Delta_\psi \,\nu_{(4D)}\Big|_- = - 2\sqrt{3}\pi\ii \Delta_\psi\frac{q}{\beta}\,.
\end{equation}
As a consequence, we have found that \eqref{eq:map5d4dblocks} holds, with
\begin{equation}
\mathcal I_N \, =\, \frac{\pi}{4}\,\Delta_\psi\, q^2\,=\,\mathcal I_S \,.
\end{equation}
The on-shell action can also be expressed in terms of the coefficients of the harmonic function as in eq.~\eqref{eq:onshellactionfinal},
\begin{equation}
I \,=\,\mathcal I_N + \mathcal I_S\,=\,  \frac{\pi}{2}\,\Delta_\psi\,q^2\,.
\end{equation}

We thus confirm that the two blocks we have identified in five dimensions match the contributions from the nut and anti-nut of the supersymmetric Killing vector in four dimensions.


\section{Multi-charge supersymmetric saddles}\label{sec:multicharge}

This section is devoted to the construction and analysis of new supersymmetric non-extremal Euclidean black hole solutions carrying multiple electric charges $Q_I$, which provide saddles of the gravitational path integral that computes a supersymmetric index refined by the conjugate electric potentials $\varphi^I$.
The solutions will be obtained by solving the supersymmetry equations directly, rather than taking a supersymmetric limit of the corresponding non-supersymmetric and non-extremal black hole as we did in section~\ref{sec:susylimit_BH} for the minimal  setup. The main reason is that non-extremal multi-charge solutions are only known for specific supergravity models~\cite{Cvetic:1996xz} or in absence of rotation \cite{Meessen:2011bd, Meessen:2012su}, while we intend to perform a general analysis.  

The section is organized as follows. We start recalling the main features of ${\cal N}=2$ supergravity coupled to vector multiplets in section~\ref{sec:matter_coupled_sugra}. Then in section \ref{sec:susy_sol_matter_coupled_sugra} we review the general classification of timelike supersymmetric solutions with an additional U(1), closely following \cite{Gutowski:2004yv, Gutowski:2004bj, Gauntlett:2004qy, Bellorin:2006yr}. In section~\ref{sec:gen_saddles} we construct our novel supersymmetric non-extremal solutions. Finally, in section~\ref{sec:physical_prop} we study their thermodynamic properties. In particular, we evaluate their on-shell action, further extending the equivariant technology of the previous sections to the matter-coupled case.

\subsection{5D supergravity coupled to vector multiplets}
\label{sec:matter_coupled_sugra}

The bosonic field content of five-dimensional $\mathcal N=2$ supergravity coupled to $n$ vector multiplets consists of the metric $g_{\mu\nu}$, $n+1$ vector fields $A^I$ and $n+1$ scalars $X^I$. The latter obey the following cubic constraint,
\begin{equation}
C_{IJK}X^I X^J X^K=1\,, 
\end{equation}
where $C_{IJK}$ is a fully-symmetric and constant tensor which determines the couplings of the bosonic action. The latter is given by
\begin{equation}\label{eq:action_sugra_matter-coupled}
S=\frac{1}{16\pi}\int \left(R\star_5 1-\frac{3}{2}\, a_{IJ}\,\diff X^I \wedge \star_5 \diff X^J-\frac{3}{2}\, a_{IJ} F^I\wedge \star_5 F^{J}
+C_{IJK}A^I\wedge F^{J}\wedge  F^K\right)\, ,
\end{equation}
where
\begin{equation}
\label{eq:defa_{IJ}&X_I}
a_{IJ}=3{X}_I {X}_{J}-2\,{C}_{IJK}X^K\,,  \hspace{1cm} X_I=C_{IJK}X^J X^K\,.
\end{equation}

In what follows we shall restrict our attention to supergravity models for which the scalar manifold is a symmetric space \cite{Gunaydin:1983bi}. This implies the existence of a fully-symmetric and constant tensor $C^{IJK}$ satisfying
\begin{equation}
C^{IJK}C_{J(LM}C_{NP)K}=\frac{1}{27} \, \delta^{I}_{(L}C_{MNP)}\, .
\end{equation}
Contracting the above equation with $X^L X^M X^N X^P$, one can invert the second of \eqref{eq:defa_{IJ}&X_I}:
\begin{equation}
X^I=27 C^{IJK}X_J X_K\, .
\end{equation}
Further contracting with $X_I$ yields
\begin{equation}
C^{IJK}X_I X_J X_K=\frac{1}{27}\, .
\end{equation}


\subsection{Timelike supersymmetric solutions with a U(1) symmetry}
\label{sec:susy_sol_matter_coupled_sugra}

Let us review the general form of supersymmetric solutions in five-dimensional $\mathcal N=2$ ungauged supergravity coupled to vector multiplets. We shall assume that the supersymmetric Killing vector is timelike. In such case, using coordinates adapted to the isometry, the metric and vector fields are given by
\begin{eqnarray}
\label{eq:susy_metric}
\diff s^2&\,=\,&-f^2\left(\diff t+\omega\right)^2 + f^{-1}\diff {\hat s}^2\,, \\[1mm]
\label{eq:susy_vector}
A^{I}&\,=\,& -X^{I} f \left(\diff t+\omega\right)+ {\hat A}^{I}+{\zeta^I}\diff t\,,
\end{eqnarray}
where $\zeta^I$ is a constant parametrizing a possible gauge choice. The metric function $f$, the scalars $X^I$, and the one-forms $\omega$ and ${\hat A}^I$ are defined on a four-dimensional hyper-Kahler (base) space whose line element is denoted by $\diff{\hat s}^2$. Supersymmetry imposes the following equations
\begin{eqnarray}
\label{eq:SUSY1}
{\hat F}^I&\,=\,&\hat{\star}\,{\hat F}^I\,, \\ [1mm]
\label{eq:SUSY2}
\diff \,{\hat\star}\, \diff \left(f^{-1}X_I\right)&\,=\,& C_{IJK} {\hat F}^{J}\wedge {\hat F}^{K}\,, \\[1mm]
\label{eq:SUSY3}
f\left(\diff \omega + {\hat \star}\,\diff \omega \right)&\,=\,& 3X_I {\hat F}^I\,, 
\end{eqnarray}
where $\hat \star$ denotes the Hodge star operator with respect to the hyper-Kahler metric.

Following the steps of section~\ref{sec:gibbonshawkingansatz}, let us assume the hyper-Kahler metric has an additional U(1) isometry respecting the triholomorphic structure. Then it must take the Gibbons-Hawking form \eqref{eq:susysolminimalghbase1}, that we repeat below for convenience,
\begin{equation}
\label{eq:basespace}
\diff {\hat s}^2=H^{-1}\left(\diff \psi +\chi\right)^2 + H \diff x^{i}\diff x^{j} \delta_{ij} \,.
\end{equation}
We recall that the function $H$ and the local one-form $\chi$ are related by  $\diff H\,=\, \star_{3} \diff \chi\,.$
Under the additional isometry, the local one-forms $\omega$, ${\hat A}^{I}$ naturally split as
\begin{eqnarray}
\omega&\,=\,&\omega_{\psi}\left(\diff \psi +\chi\right)+{\breve \omega}\,,\\[1mm]
{\hat A}^{I}&\,=\,&\frac{K^I}{H}\left(\diff \psi +\chi\right)+ {\breve A}^I\,, 
\end{eqnarray}
where again all the quantities appearing ($\omega_\psi, K^I, {\breve \omega},{\breve A}^I$) are defined on ${\mathbb R}^3$. Introducing the functions $L_I$ and $M$ as follows 
\begin{eqnarray}
f^{-1}X_I&\,=\,&L_I + C_{IJK}K^JK^K H^{-1}\,, \\[1mm]
\omega_{\psi}&\,=\,& M + \frac{3}{2} L_I K^I H^{-1}+C_{IJK} K^I K^J K^K H^{-2}\,,
\end{eqnarray}
it can then be shown that the supersymmetry equations \eqref{eq:SUSY1}--\eqref{eq:SUSY3}  boil down to the following equations
\begin{eqnarray}
\diff \star_3 \diff L_{I}&\,=\,&0\,, \hspace{1cm} \diff \star_3 \diff M\,=\,0\,\,,\\[1mm]
\label{eq:monopole}
\star_3 \diff K^{I}&\,=\,&-{\breve F}^{I}\,,\\[1mm]
\label{eq:omega3d}
\star_3\diff {\breve \omega}&\,=\,&H\diff M-M\diff H +\frac{3}{2}\left(K^I\diff L_I-L_I \diff K^I\right)\,\, .
\end{eqnarray}
Thus, in summary, timelike supersymmetric solutions with a U(1)$_{\psi}$ isometry are  determined by $2(n+2)$ harmonic functions on ${\mathbb R}^3$:
\begin{equation}
\left\{H, M, K^I, L_I\right\}\, .
\end{equation}


\subsection{Saddles for supersymmetric multi-charge black holes}
\label{sec:gen_saddles}

The aim of this section is to construct  supersymmetric solutions providing saddles of the Euclidean gravitational path integral associated to asymptotically-flat BPS (supersymmetric and extremal) black holes with electric charges $Q_I$ and angular momentum $J_{-}=\tfrac{J_1-J_2}{2}$.\footnote{The linear combination $J_{+}=\tfrac{J_1+J_2}{2}$ will vanish as a consequence of extremality.} We shall then look for Euclidean solutions ($\tau=\ii t$) with a U(1)$_{\psi} \times$ U(1)$_{\phi}$ isometry that satisfy the following asymptotic boundary conditions for $\rho\rightarrow+\infty$,
\begin{eqnarray}
\diff s^2&\,\rightarrow\,&\diff \tau^2+ \diff\rho^2 +\frac{\rho^2}{4} \left[\diff\vartheta^2+\sin^2\vartheta \,\diff \phi^2 + \left(\diff \psi+\cos\vartheta \diff\phi\right)^2\right]\,,\\[1mm]
\label{eq:bdry_cond_A^I}
A^{I}&\,\rightarrow\,& \ii \Phi^I \diff \tau\,,\\[1mm]
X^{I}&\,\rightarrow\,&\Phi^I_{*}\,,
\end{eqnarray}
along with the  twisted identifications
\begin{equation}
\label{eq:twisted_bdry_conditions}
\begin{aligned}
\left(\tau, \psi, \phi\right)\ \sim&\ \left(\tau+\beta, \psi+\beta \ii \Omega_{-}, \phi-\beta \ii \Omega_{+}\right)\,\\[1mm] 
\ \sim&\ \left(\tau, \psi+2\pi, \phi+2\pi\right)
\sim\ \left(\tau, \psi-2\pi, \phi+2\pi\right)\,\, .
\end{aligned}
\end{equation}
The last two, together with the fact that $\vartheta \in [0, \pi]$, imply that the metric asymptotes to $S^{1}_{\beta} \times {\mathbb R}^{4}$. The asymptotic behaviour of matter-coupled supersymmetric saddles we are constructing, then, falls into a subset of the ones considered around \eqref{eq:asymptoticspacetime}.

All in all, the solutions will be specified by 
\begin{equation}
\beta, \Omega_{-}, \Phi^I\hspace{1cm} \text{or, equivalently,} \hspace{1cm} Q_I, J_{+}, J_{-}\, ,
\end{equation}
and the (constrained) moduli  $\Phi^I_{*}$.  The mass $E$ of these solutions, being supersymmetric, is fixed by 
\begin{equation}
E=\Phi^I_{*} Q_I\, , 
\end{equation}
which has been shown to be equivalent to \cite{Cabo-Bizet:2018ehj, Iliesiu:2021are}
\begin{equation}
\beta \Omega_{+}=2\pi {\ii}\, .
\end{equation}

In order to construct a solution with such properties, we consider the following multi-center harmonic functions,
\begin{equation}
H=h_0 +\sum_{a} \frac{h_a}{r_a}\,, \hspace{3mm}\ii M=\mm_0 +\sum_{a} \frac{\mm_a}{r_a}\,, \hspace{3mm} \ii K^I=\kk^I_0 +\sum_{a} \frac{\kk^I_a}{r_a}\,, \hspace{3mm}L_I={\ell}_{I, 0} +\sum_{a} \frac{{\ell}_{I, a}}{r_a}\,,   
\end{equation}
where the notation is the same as in section~\ref{sec:gibbonshawkingansatz}.

Having specified the choice of harmonic functions, we can solve the equations for the one-forms $\chi$, ${\breve A}^{I}$ and ${\breve \omega}$ using two sets of spherical coordinates in $\mathbb R^3$ centered at the poles of the harmonic functions, as introduced in \eqref{eq:coordinatearoundcenters}. Locally, these are given by
\begin{eqnarray}
\label{eq:chi}
\chi&\,=\,&\left(h_N\cos \vartheta_N+h_S \cos \vartheta_S\right) \diff \phi\,, \\[1mm]
\label{eq:breve_AI}
\ii {\breve A}^{I}&\,=\,&-\left(\kk^I_N\cos \vartheta_N+\kk^I_S \cos \vartheta_S\right) \diff \phi\,, \\[1mm]
\label{eq:breve_omega}
\breve \omega&\,=\,&\left({\breve \omega}_N\cos \vartheta_N+{\breve \omega}_{S}\cos \vartheta_S\right)\diff \phi +{\breve \omega}_{\rm {reg}}\,, 
\end{eqnarray}
where 
\begin{equation}
\label{eq:omegaNS}
\begin{aligned}
\ii {\breve\omega}_N\,=\,&h_0 \mm_N-\mm_0 h_N+\frac{3}{2}\left(\kk^I_0 \ell_{I, N}-\ell_{I, 0} \kk^I_{ N}\right)-\frac{2\left(h_N\mm_S-h_S\mm_N\right)+3\left(\kk^I_N\ell_{I, S}-\kk^I_S\ell_{I, N}\right)}{2\delta}\, ,\\[1mm]
\ii {\breve\omega}_S\,=\,&h_0 \mm_S-\mm_0 h_S+\frac{3}{2}\left(\kk^I_0 \ell_{I, S}-\ell_{I, 0} \kk^I_{S}\right)+\frac{2\left(h_N\mm_S-h_S\mm_N\right)+3\left(\kk^I_N\ell_{I, S}-\kk^I_S\ell_{I, N}\right)}{2\delta}\, ,
\end{aligned}
\end{equation}
and 
\begin{equation}
\ii {\breve \omega}_{\rm {reg}}=\frac{2\left(h_N\mm_S-h_S\mm_N\right)+3\left(\kk^I_N\ell_{I, S}-\kk^I_S\ell_{I, N}\right)}{2\delta}\left(\cos\vartheta_N+1\right)\left(1-\frac{r_N+\delta}{r_S}\right)\diff \phi\, .
\end{equation}
The next step is to fix the parameters of the solution in terms of $\beta, \Phi^I, \Omega_{-}$ and the moduli by demanding asymptotic flatness and regularity. Before that, as in section~\ref{sec:gibbonshawkingansatz}, we make use of the freedom to shift $K^I\to K^{I}+ \lambda^I H$ to set
\begin{equation}
\label{eq:cond_k^I}
\kk^I_N=-\kk^{I}_S\, .
\end{equation}

\paragraph{Asymptotic flatness.} We start demanding that the metric of the base space \eqref{eq:basespace} asymptotes to ${\mathbb R}^{4}$ for large $r$, where $r$ is the radial direction of a set of coordinates centered at the origin, as given by \eqref{eq:coordinatesystem1}.\footnote{Here, $r$ is related to $\rho$ appearing in \eqref{eq:bdry_cond_A^I} by $\rho^2=4r$.} This amounts to imposing 
\begin{equation}
h_0=0\, , \hspace{1cm} h_N+h_S=1\,.
\end{equation}
Next, we impose that
\begin{equation}
\lim_{r\to \infty} f=1\,, \hspace{1cm} \lim_{r\to \infty}\omega_{\psi}=0\,,
\end{equation}
which yields the conditions 
\begin{equation}
\label{eq:asymptotic_flatness}
\kk^I_{0}=0\,, \hspace{1cm} C^{IJK}{\ell}_{I, 0}{\ell}_{J, 0}{\ell}_{K, 0}=\frac{1}{27}\,, \hspace{1cm} \mm_0=-\frac{3}{2}{\ell}_{I, 0}\left(\kk^I_N+\kk^I_S\right)=0\, ,
\end{equation}
where in the last equation we used~\eqref{eq:cond_k^I}.
\paragraph{Non-extremality and regularity conditions.} In analogy with section~\ref{sec:susyaction}, we demand a non-extremal behaviour at the centers for the metric functions $f$ and $\omega_{\psi}$. Concretely, we impose the conditions
\begin{equation}
\lim_{r_a\to 0}f^{-1}X_I= {\cal O}\left(r_a^0\right)\, , \hspace{1cm} \lim_{r_a\to 0}\omega_{\psi}= {\cal O}\left(r_a^0\right)\, ,
\end{equation}
which fix the coefficients ${\ell}_{I, a}$, $\mm_a$ in terms of $\kk^I_{a}$ and $h_a$ as
\begin{equation}
\ell_{I, a}=C_{IJK}\frac{\kk^J_a \kk^K_a}{h_a}\,, \hspace{1cm} \mm_a=-\frac{1}{2}C_{IJK}\frac{\kk^I_a \kk^J_a \kk^K_a}{h^2_a}\, .
\end{equation}
Therefore, the solution is fully determined in terms of $k^{I}_{N}, h_N$, the distance between the centers $\delta$ and ${\ell}_{I, 0}$. The latter are in correspondence with the moduli of the solution
\begin{equation}
\lim_{r\to \infty} X^{I}\,=\,\Phi^{I}_{*}\,=\, 27 \,C^{IJK}\ell_{J, 0} \ell_{K, 0}\, .
\end{equation}
Consequently, the remaining parameters must be in correspondence with the temperature and chemical potentials:
\begin{equation}
\left(\kk^{I}_{N}, h_N, \delta\right)\hspace{5mm}\leftrightarrow \hspace{5mm}\left(\beta, \Phi^I, \Omega_{-}\right)\, .
\end{equation}
The simplest way of obtaining the precise relation is by demanding regularity of the metric and gauge fields near the centers. The analysis is essentially the same as the one made in section~\ref{sec:susyaction}. The metric near the north pole has been given in \eqref{eq:metricaroundnut}. A similar expression holds around the south pole. Given these expressions, it is clear that regularity of the metric around the north and south poles imposes the conditions
\begin{equation}
\Omega_{N}=\frac{h_N}{\ii {\breve\omega}_N}\, , \hspace{1cm}\Omega_{S}=-\frac{h_S}{\ii {\breve\omega}_N}\, ,
\end{equation}
which can be explicitly verified in our solution. Now, the following change of coordinates
\begin{equation}
\psi_N=\psi+\phi -\Omega_N \tau\,, \hspace{1cm} \phi_N=\phi-\frac{\Omega_N}{2h_N}\tau\,, \hspace{1cm} {\tilde \tau}=\frac{\Omega_N}{2h_N}\tau\,, 
\end{equation}
brings the metric around the north pole \eqref{eq:metricaroundnut} to the form
\begin{equation}
\diff s^2   \, \underset{\tilde r_N\to 0}{\longrightarrow} \  \frac{f_N^2}{\Omega_N^2} \diff \psi_N^2 +\frac{h_N}{f_N}\left[d{\tilde r}_N^2+{\tilde r}_N^2\left(\frac{1}{4}\diff\vartheta_N^2+\cos^2\frac{\vartheta_N}{2}\diff {\tilde\tau}^2+\sin^2\frac{\vartheta_N}{2}\diff {\phi}^2_N\right)\right]\, .
\end{equation}
Thus, as discussed in section~\ref{sec:susyaction}, this is equivalent to the product ${\mathbb S}^1\times {\mathbb R^4}$, provided $\tilde\tau$ and $\phi_N$ are periodically identified, with period equal to $2\pi$ each. Consistency with the global identifications \eqref{eq:twisted_bdry_conditions} demands $\psi_N\sim \psi_N+4\pi$ together with
\begin{equation}
\Omega_N=-\ii \left(\Omega_{+}-\Omega_{-}\right)\,, \hspace{1cm} \Omega_S=\ii \left(\Omega_{+}+\Omega_{-}\right)\, ,\hspace{1cm} \beta=4\pi \ii{\breve \omega}_{N}\, ,
\end{equation}
which are equivalent to
\begin{equation}
\label{eq:DiracMisner}
\beta \Omega_{+}=2\pi \ii\,, \hspace{1cm} \beta \Omega_{-}=2\pi \ii\left(h_S-h_N\right)\,, \hspace{1cm} \beta=4\pi \ii{\breve \omega}_{N}\, .
\end{equation}
An alternative way of arriving to the same result would be to study the conditions for the cancellation of Dirac-Misner singularities. This is discussed in Appendix~\ref{sec:DiracMisner}.

Finally, we obtain the relation between the parameters $\kk^I $ and the chemical potentials $\Phi^I$. To this aim, we first demand regularity of the gauge fields at the north and south poles. Namely, we impose $\iota_{\xi_N}A^I\big|_{{\cal M}_N}=\iota_{\xi_S}A^I\big|_{{\cal M}_S}=0$. Both yield the same condition, which is the following
\begin{equation}
\label{eq:reg_AI}
\beta {\zeta}^{I}=-4\pi \kk_N^I\, .
\end{equation}
Now we make use of the boundary conditions \eqref{eq:bdry_cond_A^I} to obtain $-\zeta^I+\Phi^I_{*}=\Phi^I$. This allows us to rewrite the above regularity condition \eqref{eq:reg_AI} as follows
\begin{equation}\label{eq:PhiI~kI}
\beta(\Phi^I-\Phi^I_{*})\,=\,4\pi \kk_N^I\, .
\end{equation}

\paragraph{Moduli independence.} 
 Contrarily to what happens in the supersymmetric extremal case, where the values of the scalar fields at the horizon are completely fixed by the charges of the solution (see e.g.\ the review~\cite{Larsen:2006xm}), we find that the scalar fields at the centers of the harmonic functions in our finite-temperature setup depend on their values at infinity (the moduli) as well as all other parameters of the solution. However, we observe that there is a combination of the fields that has a much simpler dependence on the parameters. 
 This is the combination that gives the scalars of four-dimensional $\mathcal{N}=2$ supergravity upon dimensional reduction along the $\partial_\psi$ direction, that is (see e.g.~\cite{Behrndt:2005he})
\be
A^I_\psi   \pm \ii \sqrt{g_{\psi\psi}}\, X^I\,.
\ee
While in Lorentzian signature the two sign choices in this expression yield complex conjugate quantities,  in our Euclidean context $A_\psi$ is purely imaginary, hence both sign choices give purely imaginary (and a priori independent) quantities.

Let us illustrate the claim above in some detail. From the formulae given above in this section, we have
\be
A^I_\psi   \pm \ii \sqrt{g_{\psi\psi}}\, X^I \,=\,  \frac{K^I}{H} + \ii X^I f(\ii\omega_\psi) \pm \ii  X^I\sqrt{f^2(\ii\omega_\psi)^2+(fH)^{-1}}\,,
\ee
where we recall that our reality conditions are such that $\ii K^I$ and $\ii \omega_\psi$ are real. As we approach one of the poles by taking $r_a\to 0$, we have that $(fH)^{-1}\to 0$, while $\ii\omega_\psi \to -1/\Omega_a$ hence
\be
\lim_{r_a\to 0} \left(A^I_\psi   \pm \ii \sqrt{g_{\psi\psi}}\, X^I \right)  \ = \  - \ii \,\frac{\kk^I_a}{h_a} - \ii \,\frac{fX^I|_a}{\Omega_a}\left(1 \mp  {\rm sign}(\Omega_a)  \right)\,.
\ee
Depending on the sign of $\Omega_a$, the second term cancels out for one of the two sign choices. In this case, we are left with 
\be
\lim_{r_a\to 0} \left(A^I_\psi   \pm \ii \sqrt{g_{\psi\psi}}\, X^I \right)  \ = \  - \ii \,\frac{\kk^I_a}{h_a} \,.
\ee
which is independent both of the asymptotic values of the scalars and the distance between the centers $\delta$. The result can naturally be expressed in terms of the supersymmetric chemical potentials to be introduced in \eqref{eq:chemical_pot_multicharge} below.
These findings for our five-dimensional Euclidean black hole saddles are in line with those for four-dimensional black hole saddles discussed in~\cite{Boruch:2023gfn}.


\subsection{Physical properties of the solutions}\label{sec:physical_prop}

In this section we first make use of the equivariant approach developed in section \ref{sec:equiv5d} to compute the on-shell action of the Euclidean solutions derived in section \ref{sec:gen_saddles}. 
We then perform its Legendre transform to obtain the entropy, allowing at the same time for more general values of the parameters through analytic continuations. Even assuming the charges are real, the entropy we obtain through this procedure is in general complex. Demanding that the imaginary part vanishes yields a constraint on the charges of the solution. We find that there are two possibilities, depending on the assumptions  on the charges. The first possibility, which gives a finite entropy, corresponds to the extremal limit $\beta\to \infty$ of our configurations. We explicitly show that this extremal limit gives known BPS black hole solutions \cite{Breckenridge:1996is, Cvetic:1996xz, Chamseddine:1998yv, Gutowski:2004bj, Ortin:2015hya} after Wick-rotating to Lorentzian time, analogously to other similar cases discussed in previous work.  The second possibility is a new observation: we find that it corresponds to the limit $\beta \to 0$ and yields a supersymmetric topological soliton with vanishing entropy. In this case, we show that the corresponding Lorentzian solution is a two-center horizonless microstate geometry~\cite{Giusto:2004id,Bena:2005va, Berglund:2005vb}. Therefore, both the BPS black hole and the horizonless geometry arise as limits of our supersymmetric, finite $\beta$ solution.\footnote{Although we illustrate these results  in the present multi-charge setup, we remark that they also apply to the pure supergravity case of section~\ref{sec:susyaction}, consistently with the discussion in section~\ref{contribindex}.}


\subsubsection{Euclidean on-shell action via equivariant integration}
\label{sec:onshellaction_multicharge}

After using the trace of Einstein equations, we can recast the bulk contribution to the on-shell action as 
\begin{equation}
I_{\rm{bulk}}=\frac{1}{16\pi}\int F^I\wedge G_I\,,
\end{equation}
where 
\begin{equation}
G_{I}=a_{IJ}\star_5 F^J-\ii \, C_{IJK}F^J\wedge A^K\, .
\end{equation}
We emphasize that $\star_5$ is the Hodge star operator with respect to the Euclidean metric, hence the appearance of the imaginary unit $\ii$. The matter-coupled generalization of the equivariantly-closed polyform in \eqref{eq:polyform5d2} is
\begin{equation}
\Psi=F^I\wedge G_I + \left[-\left(\iota_{\xi}A^I+\crho^I\right)G_I+\nu_{I}\wedge F^I \right]+ \left[-\left(\iota_{\xi}A^I+\crho^I\right) \nu_I\right]\,, 
\end{equation}
where $\crho^I$ are for the time being arbitrary constants and $\nu_I$ is defined as follows
\begin{equation}
\diff {\nu}_{I}=\iota_{\xi}G_I\,, \hspace{1.5cm} \iota_{\xi}\nu_I=0\,.
\end{equation} 

As in section~\ref{sec:susyaction}, we restrict to supersymmetric solutions with an extra U(1)$_{\psi}$ isometry, and consider two linear combinations of the supersymmetric Killing vector and the generator of the extra U(1) isometry, namely
\begin{equation}
\xi_{a}=\partial_{\tau}+ \Omega_{a} \, \partial_{\psi}\, ,
\end{equation}
where
\begin{equation}
\Omega_{a}=-\frac{1}{\ii \omega_{\psi}}\Big|_{{\cal M}_a}\, , \hspace{1cm} a=N, S\, .
\end{equation}
We choose to localize with respect to $\xi_N$. The determination of the equivariant parameters follows exactly the same steps as in section~\ref{sec:susyaction}. Therefore, we just provide the final result
 \begin{equation}
\label{eq:equivariant_parameters}
\epsilon^{N}_1=\frac{2\pi}{\beta}=-\epsilon^{N}_2\,.
\end{equation}

The one-form $\nu_I$ associated to $\xi_{N}$ is given by 
\begin{equation}
\begin{aligned}
\nu_I\,&=\, \ii C_{IJK}\left[\left(f X^J+\zeta^J\right)\ii {\hat A}^{K}-fX^J \zeta^{K} \ii \omega\right] - \ii \Omega_{N} \Big[\varepsilon_{I}+C_{IJK}fX^JH^{-1}\ii K^K \left(\diff\tau+\ii {\breve \omega}\right) \\[1mm]
&\quad\ \left.+\,C_{IJK}\left(H^{-1}\ii K^J-f X^J\ii\omega_{\psi}\right)\left(\ii {\breve A}^{K}+\zeta^K \diff \tau\right)\right]+ \ii\mu_{I}\, ,
\end{aligned}
\end{equation}
where the one-form $\varepsilon_I$ is related to the function $L_I$ through $\diff \varepsilon_I=\star_3 \diff L_I$, and $\mu_I$ is a closed one-form to be determined demanding regularity of $\nu_I$. Let us emphasize that except for this regularity analysis, which must be done case by case, the above expression for $\nu_I$ applies to any timelike supersymmetric solution with an extra U(1)${}_{\psi}$. As we saw in the previous sections, there is a scheme, which here corresponds to choosing
\begin{equation}\label{eq:choice_cI}
\ii \crho^{I}\,=\, \Phi^I-\Phi^I_{*}\, ,
\end{equation}
 such that the boundary terms cancel and the whole on-shell action comes from the integral of $\nu_I$ at the nut.

In what follows we specialize to the solutions studied in section \ref{sec:gen_saddles}. Demanding regularity, we obtain that 
\begin{equation}
\mu_I=C_{IJK}\,\zeta^{J}\,\frac{\kk^K_N}{h_S}\left(\diff \psi-\Omega_N \diff \tau\right)\, ,
\end{equation}
which already allows us to evaluate the nut contribution to the on-shell action. We will also assume the choice \eqref{eq:choice_cI}. Therefore the on-shell action is given by
\begin{equation}
\label{eq:onshellaction_multicharge}
I\,=\,\frac{1}{16\pi}\int_{{\cal M}_{N}} \frac{\iota^* \Psi_{(1)}}{\frac{\epsilon^N_1}{2\pi} \frac{\epsilon^N_2}{2\pi}}\,=\,-4\pi^2\frac{C_{IJK}\kk^I_N\kk^J_N\kk^K_N}{h_N h_S}\,=\,\pi \,\frac{C_{IJK}\varphi^I \varphi^J\varphi^K}{\omega_{+}^2-\omega_{-}^2}\, ,
\end{equation}
where we have introduced the quantities $\varphi^I, \omega_{+}, \omega_{-}$, defined as  
\begin{equation}
\varphi^I=\beta\left(\Phi^I-\Phi^I_{*}\right)\,, \hspace{1cm} \omega_{+}=\beta \Omega_{+}\,, \hspace{1cm} \omega_{-}=\beta \Omega_{-}\, ,
\end{equation}
which in terms of the parameters of the solutions read
\begin{equation}
\label{eq:chemical_pot_multicharge}
\varphi^I=4\pi \kk^I_N\,, \hspace{1cm} \omega_{+}=2\pi \ii \hspace{1cm} \omega_{-}=2\pi \ii \left(h_S-h_N\right)\, .
\end{equation}


\subsubsection{Entropy and charges from the on-shell action} 

The entropy ${\cal S}$ of the saddles follows from the extremization principle:
\begin{eqnarray}
\label{eq:S}
{\cal S}&=&{\rm{ext}}_{\{\varphi^I, \omega_{\pm}, \Lambda\}}\left[-I-\varphi^I Q_I -\omega_{-} J_{-}-\omega_{+} J_{+}-\Lambda\left(\omega_{+}-2\pi \ii\right)\right]\, ,\\[1mm]
Q_I&=& -\frac{\partial I}{\partial \varphi^I}\,, \hspace{1cm} J_{-}=-\frac{\partial I}{\partial \omega_{-}}\,, \hspace{1cm} J_{+}=-\frac{\partial I}{\partial \omega_{+}}-\Lambda\,, \hspace{1cm} \omega_{+}=2\pi \ii\, ,\quad
\end{eqnarray}
which gives rise to the following expression (see e.g.~\cite{Cassani:2024tvk} for additional details)
\be\label{eq:entropy_multicharge}
{\cal S}\,=\,4\sqrt{\pi}\sqrt{C^{IJK}Q_I Q_J Q_K-\frac{\pi}{4}J_{-}^2}-2\pi \ii J_{+} \,.
\ee
The expressions for the charges $Q_I$ and the angular momentum $J_{-}$ as a function of the parameters of the solution can be obtained using the above extremization equations, together with \eqref{eq:onshellaction_multicharge} and \eqref{eq:chemical_pot_multicharge}:
\begin{equation}
\label{eq:charges}
Q_{I}=3\pi\frac{C_{IJK}\kk^J_N \kk^K_N}{h_N h_S}\,, \hspace{1cm}  \ii  J_{-}=\pi \frac{h_S-h_N}{h_N^2 h_S^2}C_{IJK}\kk^I_{N}\kk^J_N \kk^K_N\, .
\end{equation}
In turn, a direct calculation shows that $J_{+}$ is  given by
\begin{equation}
\label{eq:Jphi}
\ii J_{+}\,=\, 3\pi \ell_{I, 0}\kk^I_N \delta\, .
\end{equation}

Now we would like to interpret the entropy \eqref{eq:entropy_multicharge} and the charges \eqref{eq:charges} and \eqref{eq:Jphi} as those corresponding to a physical (Lorentzian) solution. To this aim, we assume first that there exists an analytic continuation of the parameters such that the charges and angular momenta become real.  Even after assuming this, the entropy \eqref{eq:entropy_multicharge} remains complex unless a constraint is imposed on the charges.
The constraint that one must impose clearly depends on the sign of the combination of the charges appearing in the argument of the square root, 
\be\label{eq:argument_squareroot}
C^{IJK}Q_I Q_J Q_K-\frac{\pi}{4}J_{-}^2\, .
\ee
When this is positive, the entropy is real only when $J_{+}=0$. It is well known that this is the constraint satisfied by a supersymmetric and extremal black hole with these charges \cite{Breckenridge:1996is, Cvetic:1996xz, Chamseddine:1998yv, Gutowski:2004bj, Ortin:2015hya}. Therefore, $J_{+}=0$ must follow from extremality $\beta\to \infty$. It is instructive to explicitly verify this by expressing $\beta$ as a function of the charges. Using the above expressions for the charges and angular momenta in terms of the parameters, alongside with \eqref{eq:DiracMisner} and \eqref{eq:omegaNS}, one finds that 
\begin{equation}
\label{eq:beta(Q)}
\beta=-18\,\frac{C^{IJK}{\bar X}_I Q_J Q_K}{{\cal S}+2\pi \ii J_{+}}\frac{\cal S}{\ii J_{+}}\, ,
\end{equation}
where ${\bar X}_{I}\equiv C_{IJK}\Phi^{J}_{*}\Phi^K_{*}$. Thus, we conclude that an extremal solution with finite entropy must necessarily have vanishing $J_{+}$. The Lorentzian black hole solution with these properties is studied in section \ref{label:extBHs}, where we show that its entropy agrees with \eqref{eq:entropy_multicharge}.

The second possibility is that \eqref{eq:argument_squareroot} is non-positive.\footnote{The limit in which \eqref{eq:argument_squareroot} vanishes can also be reached from the extremal black hole, but it gives rise to a singular solution, as it corresponds to a black hole with vanishing horizon area. This is the reason why we shall not discuss this possibility.} In such case the expression \eqref{eq:entropy_multicharge} is purely imaginary, implying that it should vanish. This is equivalent to the following relation among the charges,
\begin{equation}
\label{eq:constraint_horizonless_sol}
C^{IJK}Q_I Q_J Q_K-\frac{\pi}{4}\left(J_{-}^2- J^2_{+}\right)=0\,,
\end{equation}
which is then expected to be realized in a horizonless solution. From \eqref{eq:beta(Q)} it is further deduced that such solution must arise in the $\beta \to 0$ limit, provided the charges are not vanishing. Interestingly, we find that the Lorentzian solution that realizes these properties is the two-center horizonless geometry of \cite{Bena:2005va, Berglund:2005vb}. This is further discussed in section \ref{sec:fuzz}. 


\subsubsection{$\beta\to \infty$ limit: BPS black holes}
 \label{label:extBHs}
From the last of \eqref{eq:DiracMisner} and the expression for ${\breve \omega}_{N}$ provided in \eqref{eq:omegaNS}, we see that the $\beta\to\infty$ limit corresponds to the limit in which the distance between the centers goes to zero. Taking the $\delta\to 0$ limit in the harmonic functions yields
\begin{equation}
H \to \frac{1}{r}\,, \hspace{1cm} K^{I}\to 0\,, \hspace{1cm} L_I \to \ell_{I, 0}+\frac{\ell_I}{r}\,, \hspace{1cm} M \to \frac{m}{r}\,,
\end{equation}
where
\begin{equation}
\ell_I=C_{IJK}\frac{\kk^J_N \kk^{K}_N}{h_Nh_S}\,, \hspace{1cm} \ii m=\frac{1}{2}C_{IJK}\kk^{I}_N \kk^J_N \kk^{K}_N \frac{ \left(h_N-h_S\right)}{h_N^2h_S^2}\, .
\end{equation}
After both a Wick rotation and an analytic continuation of $h_N-h_S$,
\begin{equation}
\label{eq:Wick}
\tau\to \ii t\,, \hspace{1cm} h_N\to\frac{1}{2}\left(1+\ii \Delta_h\right)\,, \hspace{1cm} h_S\to\frac{1}{2}\left(1-\ii \Delta_h\right)\,,
\end{equation}
one obtains a BPS (supersymmetric and extremal) black hole with charges $Q_I$ and angular momentum $J_{-}$ given by
\begin{equation}
\label{eq:charges_extremal}
Q_I=3\pi \ell_I=6\pi \,\frac{C_{IJK}\kk^J_N \kk^K_N}{1+\Delta^2_h}\,,  \hspace{1cm} J_{-}=-2\pi m=-4\pi \frac{\Delta_h\, C_{IJK}\kk^I_N \kk^J_N \kk^K_N}{\left(1+\Delta^2_h\right)^2} \, ,
\end{equation}
while $J_{+}$ vanishes. The Bekenstein-Hawking entropy, computed as one quarter of the horizon area, is
\begin{equation}
\label{eq:Sextremal}
{\cal S}=4\sqrt{\pi}\sqrt{C^{IJK}Q_I Q_J Q_K-\frac{\pi}{4}J_{-}^2}\, ,
\end{equation}
in precise agreement with the Legendre transform of the saddle on-shell action \eqref{eq:S}, supplemented with the reality condition $J_{+}=0$.

\subsubsection{$\beta\to 0$ limit: two-center microstate geometries}
\label{sec:fuzz}
Finally we discuss the $\beta\to 0$ limit. Using again \eqref{eq:DiracMisner} and \eqref{eq:omegaNS}, we get that the condition $\beta \to 0$ fixes the distance between the centers in terms of the remaining parameters,
\begin{equation}
\label{eq:bubble_eq}
\delta= - \frac{C_{IJK}\kk^I_N \kk^J_N \kk^K_N}{3 \ell_{I, 0} \kk^I_N h_N^2 h_S^2}\, .
\end{equation}
In this limit, a real Lorentzian solution is obtained by the analytic continuation 
\begin{equation}
\label{eq:analytic_cont_microstate_geom}
\tau\to \ii t\,, \hspace{1cm} \kk^I_{N}\to \ii k^I_N\,.
\end{equation}
In particular one can verify that the charges $Q_{I}$ and both angular momenta $J_{\pm}$ become real, satisfying \eqref{eq:constraint_horizonless_sol}, which implies a vanishing entropy. The solution obtained following this procedure corresponds to a horizonless two-center microstate geometry \cite{Bena:2005va, Berglund:2005vb}. Indeed, the relations satisfied by the coefficients of the harmonic functions in microstate geometries are identical to those imposed for the saddles in section~\ref{sec:gen_saddles}. The main difference between these two solutions lies on global aspects. More concretely, the conditions arising from the removal of Dirac-Misner singularities are different, ultimately due to the fact that in our saddles the time coordinate is periodically identified. For microstate geometries, the absence of Dirac-Misner strings imposes two conditions \cite{Bena:2005va, Berglund:2005vb}: the so-called `bubble equations' (thoroughly studied in \cite{Avila:2017pwi}) and that $h_N$ is an integer. The bubble equation\footnote{There are $s-1$ bubble equations, being $s$ the number of centers. } is precisely recovered upon taking the $\beta\to 0$ limit of the last of \eqref{eq:DiracMisner}. As a matter of fact, one can check that the analytic continuation of \eqref{eq:bubble_eq} precisely corresponds to the solution to the bubble equation. In turn, the condition $h_N\in {\mathbb Z}$ follows from demanding consistency of the twisted identifications \eqref{eq:twisted_bdry_conditions} when $\beta$ is set to zero. Indeed, using \eqref{eq:DiracMisner} we see that they imply $\left(\psi, \phi\right)\sim \left(\psi +2\pi \left(h_N-h_S\right), \phi+2\pi\right)\sim \left(\psi +2\pi, \phi+2\pi\right)\sim \left(\psi -2\pi, \phi+2\pi\right)$, which only makes sense when $h_N$ (and therefore $h_S=1-h_N$) is an integer.

\section{Discussion}\label{sec:discussion}
  
In this paper, we have studied equivariant localization of the ungauged supergravity on-shell action in five dimensions, with or without supersymmetry. We have  shown that there is  a scheme choice such that the boundary terms cancel out, and one is just left with nut and bolt contributions. While in general this choice depends on the conserved charges of the solution,  if we assume supersymmetry then it only depends on boundary conditions and thus acquires universal value.
 
We have examined in detail the features that are specific to odd dimensions and thus did not appear in previous even-dimensional investigations: the nuts are one-dimensional loci rather than isolated points; their contribution to the action is given by the holonomy of a one-form potential, denoted by $\nu$; global regularity of $\nu$ gives non-trivial conditions, as the fixed locus at the nut may collapse elsewhere on the manifold. 
In particular, we have seen that in black hole solutions with $S^3$ horizon topology and biaxial  ${\rm U}(1)^2$ symmetry, one can combine the Killing vector generating the thermal circle with either one of the axial Killing vectors, obtaining two independent vectors, each with a single nut (generated by the other axial ${\rm U}(1)$). In this context, we have found that the equivariant localization formula for the action using either one of these vectors splits into a sum of two terms, with the second term being naturally associated with the nut of the other vector. The final expression reminds of a sum over gravitational blocks~\cite{Hosseini:2019iad}. This is particularly evident in the supersymmetric case, where the two terms appearing in the final expression can be seen as the contribution from the two centers of the harmonic functions that are used to build the supersymmetric solution.

The latter observation suggests that there should be a generalization of our result to supersymmetric solutions with a Gibbons-Hawking base-space where the harmonic functions have more than two centers. This would give an action formula encompassing more black objects, such as multi-center black holes, black lenses \cite{Kunduri:2014kja} and black rings \cite{Elvang:2004rt,Gauntlett:2004wh,Gauntlett:2004qy}, 
as well as the solutions discussed in~\cite{Katona:2022bjp}. We plan to report on this generalization in the near future. A further development would be to extend our approach to more general solutions possessing an axial ${\rm U}(1)$ symmetry that does not preserve the Killing spinor, such as those of \cite{Bena:2007ju}.
 

Besides illustrating the `nuts and bolts' of equivariant localization in five-dimensional supergravity, we have discussed its application to the evaluation of the action of supersymmetric non-extremal black hole solutions, which represent saddle-point contributions to the gravitational path integral computing a supersymmetric index. We have constructed multi-charge, doubly-rotating such solutions in supergravity coupled to vector multiplets, and demonstrated  global regularity of the Euclidean section. Then, we have computed their action equivariantly and discussed their supersymmetric thermodynamics. While the supersymmetric action is independent of the inverse temperature $\beta$, the solutions do depend on it. By tuning $\beta$ we have found that the black hole saddles interpolate between two limiting solutions, both having a Lorentzian interpretation: the $\beta\to \infty$ limit gives the extremal supersymmetric black hole, while by taking $\beta\to 0 $ we obtain solitonic solutions carrying a vanishing Bekenstein-Hawking entropy, which are identified with two-center microstate geometries. The on-shell action takes the same functional form in terms of the supersymmetric chemical potentials 
 $\omega_1,\omega_2,\varphi^I$, however $\omega_1-\omega_2$ and $\varphi^I$ take different values in the two solutions as a consequence of different analytic continuations of the parameters being implemented. We have shown how the Legendre transform of the action encompasses both solutions by imposing different conditions on the charges, which in the solitonic case lead to a vanishing entropy. It would be interesting to clarify this intriguing connection in the context of black hole microstate counting and explore it more widely, for instance by comparing the on-shell actions in different regimes of the chemical potentials with a microscopic computation of the index. 

One can also investigate the features above in asymptotically AdS solutions to gauged supergravity.  Consider for instance minimal gauged supergravity. In this case, the finite-temperature supersymmetric saddles were discussed in~\cite{Cabo-Bizet:2018ehj}, and the two limiting Lorentzian solutions are just the supersymmetric extremal black hole and the supersymmetric topological soliton found in~\cite{Chong:2005hr}.

There are further directions for future work that it would be interesting to pursue. 
A first one is to develop equivariant localization for the Euclidean on-shell action of general asymptotically AdS solutions to five-dimensional gauged supergravity, which would allow to compare with the partition function of  the holographically dual  field theories in general curved backgrounds. We expect that the main features discussed in the present paper for the ungauged case will also hold in the presence of a cosmological constant, however the technical details may be different. 

One could also study equivariant localization of the on-shell action in odd dimensions other than five, as well as the application to the evaluation of  volumes of compactification manifolds~\cite{Martelli:2006yb}. 

Finally, it would be interesting to apply equivariant localization techniques to off-shell and higher-derivative supergravity. While finding the higher-derivative corrections to two-derivative solutions is in general very hard, use of equivariant localization may allow to bypass this step and evaluate the higher-derivative on-shell action  by knowing just minimal information about the solution, that may possibly be fixed at the two derivative level. 
 Application to supersymmetric non-extremal black hole saddles may allow to extend the recent results of~\cite{Chowdhury:2024ngg,Chen:2024gmc,Cassani:2024tvk} for asymptotically flat black holes, and~\cite{Bobev:2022bjm,Cassani:2022lrk,Cassani:2024tvk} for asymptotically AdS black holes, beyond first-order in the corrections.

\section*{Acknowledgments}

We acknowledge useful discussions with Pietro Benetti Genolini, Massimo Bianchi, Nikolay Bobev, Stefano Giusto, Kiril Hristov, Dario Martelli, Sameer Murthy, Antonio Pittelli,  Pedro F.~Ram\'irez,  Davide Rovere, Dan Waldram, and Alberto Zaffaroni. DC is supported in part by the MUR-PRIN contract 2022YZ5BA2 - Effective quantum gravity. AR is supported
by a postdoctoral fellowship associated to the MIUR-PRIN contract 2020KR4KN2 - String
Theory as a bridge between Gauge Theories and Quantum Gravity and in part by the INFN Sezione di Roma Tor Vergata. AR further
thanks the High Energy Theory group at INFN and University of Padova for hospitality.

\appendix

\section{Localization for the 4D Euclidean black hole saddle}\label{sec:4dexample}

Equivariant localization of the on-shell action of four-dimensional (gauged) supergravity has been studied in~\cite{BenettiGenolini:2023kxp,BenettiGenolini:2024xeo,BenettiGenolini:2019jdz}. In this appendix, we provide a non-supersymmetric discussion in the ungauged case, apply it to the Euclidean Kerr-Newman solution and take its supersymmetric limit at the end. The results are used in section~\ref{sec:exampleaction4d}.

\subsection{Einstein-Maxwell on-shell action as an equivariant integral}

We consider Einstein-Maxwell theory in four dimensions, which is also the bosonic sector of pure $\mathcal{N}=2$ supergravity, and show that the Euclidean on-shell action of solutions admitting a U(1) symmetry can be computed equivariantly. 

Using the trace of the Einstein equation, the bulk contribution to the on-shell action can be written as\footnote{We take a non-canonical normalization for the four-dimensional gauge field  in order to align with the conventions used in section~\ref{sec:exampleaction4d}.}
\begin{equation}
\label{eq:EM}
I_{\rm bulk}\,=\,-\frac{1}{16\pi G_4}\int_{{\cal M}} \left(R\star_4 1-\frac{2}{3}F\wedge\star_4 F\right)\,=\,\frac{1}{24\pi G_4}\int_{{\cal M}} F\wedge \star F\, ,
\end{equation}
which equals the integral on ${\cal M}$ of the following $\xi$-equivariantly-closed polyform
\begin{equation}
\label{eq:polyform_EM}
{\Psi}\,=\,F\wedge \star_4 F\,+\,\left(-\left(\iota_{\xi}A+c\right)\star_4 F+\nu\, F\right)\,+\,\left(-\left(\iota_{\xi}A +c\right) \nu\right)\,,\end{equation}
where $\xi$ is the Killing vector field generating the ${\rm U}(1)$ symmetry and  $\nu$ is the zero-form satisfying the differential equation
\begin{equation}
\diff{\nu}=\iota_{\xi}\star_4 F\,.
\end{equation}
This implies ${\cal L}_\xi \nu = 0$, and we have assumed a gauge such that ${\cal L}_{\xi}A=0$, so that ${\cal L}_\xi \Psi =0$.

We can thus evaluate $I_{\rm bulk}$ using the equivariant localization formula~\eqref{eq:BVAB_bdry}. 
The full on-shell action is
\begin{equation}
I = I_{\rm bulk} + I_{\rm GHY}\,,
\end{equation}
where, as in the main text, the GHY term is
\begin{equation}
I_{\rm GHY} = -\frac{1}{8\pi G_4}\int_{\partial {\cal M}} \diff^3 x\left( \sqrt{h}\, \mathcal K - \sqrt{h_{\rm bkg}}\,\mathcal K_{\rm bkg}\right)\,.
\end{equation}

We can then use the localization formula~\eqref{eq:BVAB_bdry} to compute the on-shell action of solutions with a U(1) symmetry. We illustrate this with the Kerr-Newman solution.

\subsection{On-shell action of Kerr-Newman} \label{sec:KNaction} 

The Kerr-Newman solution (with no magnetic charge) in Euclidean signature reads
\begin{eqnarray}
\label{eq:KN4dsol}
\diff s^2&=&\frac{\Delta_r}{r^2-\aa^2\cos^2 \vartheta}\left(\diff \tau+\aa \sin^2\vartheta\diff \phi\right)^2+ \left(r^2-\aa^2\cos^2 \vartheta\right)\left(\frac{\diff r^2}{\Delta_r}+\diff \vartheta^2\right)\quad\\[1mm]
&&+\,\frac{\sin^2\vartheta}{r^2-\aa^2\cos^2 \vartheta}\left(\left(r^2-\aa^2\right)\diff\phi-\aa\diff \tau\right)^2\,, \nonumber\\[1mm]
\ii A&=&\frac{\sqrt{3}\,q\,r}{r^2-\aa^2\cos^2 \vartheta}\left(\diff \tau+\aa\sin^2\vartheta \diff \phi\right)-\Phi\, \diff \tau\,,
\end{eqnarray}
where $\Phi$ is the electrostatic potential, $\Delta_r=\left(r-r_+\right)\left(r-r_{-}\right)$, and $r_{\pm}=m \pm \sqrt{m^2+\aa^2-q^2}$.
We demand $r_+^2>\aa^2$, so that the metric is positive-definite for $r>r_+$.
 The parameter $m$ is directly identified with the ADM mass, namely $E = m/G_4$, while the electric charge reads $Q=q/(\sqrt{3}G_4)$ and the angular momentum is $J=\ii \aa E$. The expressions for the inverse temperature $\beta$, the electrostatic potential $\Phi$ and the angular velocity $\Omega$ are
\begin{equation}
\beta=\frac{4\pi\left(r_+^2-\aa^2\right)}{r_+-r_-}\,, \hspace{1.5cm} \Phi= \frac{\sqrt{3}\,q\,r_+}{r_+^2-\aa^2}\,, \hspace{1.5cm} \Omega=\frac{\ii \,\aa}{r_+^2-\aa^2}\, . 
\end{equation}

We consider the Killing vector,
\begin{equation}\label{eq:localizing_vector_4d}
\xi=\partial_{\tau}\, ,
\end{equation}
which has a nut at the north pole of the horizon ($r=r_+, \vartheta=0$) and an anti-nut at the south pole ($r=r_+, \vartheta=\pi$)~\cite{Gibbons:1979xm}. The equivariant parameters are given by
\begin{equation}
\epsilon^{\pm}_{1}=\frac{2\pi}{\beta}\,, \hspace{1cm} \epsilon^{\pm}_{2}=\pm \ii \Omega\, ,
\end{equation}
where the $\pm$ refers to the nut ($+$) and anti-nut ($-$).  All the information required to compute \eqref{eq:polyform_EM} is encoded in $\iota_\xi A$ and $\nu$. An explicit computation yields
\begin{equation}\label{eq:iotaAandnu4d}
\iota_{\xi}A \,=\,-\frac{\ii \sqrt{3} \,q\, r}{r^2-\aa^2\cos^2\vartheta}+\ii \Phi\,, \hspace{1.5cm}
\nu\,=\,-\frac{\ii\sqrt{3}\, \aa \,q\cos\vartheta}{r^2-\aa^2 \cos^2\vartheta}\, .
\end{equation}

We can now evaluate the action. From the boundary terms in~\eqref{eq:BVAB_bdry} we get:
\begin{equation}
-\frac{1}{24\pi G_4}\int \eta \wedge \Psi_{(2)}\,=\, \frac{\beta Q}{2}\left( -\Phi + \ii c\right) \, ,
\end{equation}
whereas the boundary term involving $\Psi_{(0)}$ vanishes. From the fixed points we obtain the contribution
\begin{equation}
\label{eq:nut_contr_KN}
\frac{(2\pi)^2}{24\pi G_4} \left(\frac{\Psi_{(0)}|_{+}}{\epsilon_1^+\, \epsilon_2^+}+\frac{\Psi_{(0)}|_{-}}{\epsilon_1^- \,\epsilon_2^-}\right)\,=\,  -\ii c\,\frac{\beta Q}{2}\, .
\end{equation}
Putting together the contributions to the equivariant integral, we obtain
\begin{equation}
\frac{1}{24\pi G_4}\int_{\cal M}\Psi\,=\, -\frac{\beta \Phi Q}{2}\, ,
\end{equation}
that is independent of $c$.
Finally, the contribution from the GHY term is
\begin{equation}
I_{\rm GHY}\,=\,\frac{\beta E}{2}\, .
\end{equation}
Thus, we arrive at the final expression for the on-shell action
\begin{equation}
I\,=\,\frac\beta2\left(E-\Phi Q\right)\, .
\end{equation}
We note that when fixing the constant $c$ as
\begin{equation}
\label{eq:c4d}
\ii c\,=\, \Phi - \frac{E}{Q}\, ,
\end{equation}
the contribution to the on-shell action coming from the boundary terms vanishes, and we are just left with the contribution \eqref{eq:nut_contr_KN} from the fixed points of the isometry:
\begin{equation}\label{eq:susyaction4dsum}
I\,=\,\frac{(2\pi)^2}{24\pi G_4}\left(\frac{\Psi_{(0)}|_{+}}{\epsilon^{+}_{1}\, \epsilon^{+}_ {2}}+\frac{\Psi_{(0)}|_{-}}{\epsilon^{-}_{1}\, \epsilon^{-}_ {2}}\right)\,=\,-\frac{\beta}{2}  \left(\Phi-E/Q\right)Q\, =\,\frac{\beta}{2}\left(E-\Phi Q\right)\, .
\end{equation}

We now consider the supersymmetric (non-extremal) limit. This amounts to taking
\begin{equation}
E=\sqrt{3}Q\, ,
\end{equation}
or equivalently $m=q$, as dictated by the supersymmetry algebra. In terms of the potentials, this corresponds to fixing $\beta\Omega = \pm 2\pi \ii$. 
The special value of $\crho$ given by \eqref{eq:c4d} is fixed to $\crho = -\ii\left( \Phi - \sqrt{3}\right)$, which depends on the electrostatic potential and thus on boundary conditions, only. We also note that the Killing vector \eqref{eq:localizing_vector_4d} used to localize coincides with the one arising as a bilinear of the Killing spinor.
The supersymmetric limit of the on-shell action \eqref{eq:susyaction4dsum}  reads in terms of the parameters,
\begin{equation}\label{eq:susyaction4d}
I \,=\, \frac{\pi q^2}{G_4} \,.
\end{equation}
It can be conveniently written in terms of the supersymmetric chemical potential $\varphi$, defined as
\begin{equation}
\varphi = \beta\big( \Phi - \sqrt{3}\,\big) = -6\pi G_4Q \,,
\end{equation}
which gives
\begin{equation}
I \,=\, \frac{\varphi^2}{12\pi G_4}\,.
\end{equation}
Using this expression, one can study the supersymmetric thermodynamics of the solution~\cite{Iliesiu:2021are,Hristov:2022pmo}.

\section{Absence of Dirac-Misner strings}
\label{sec:DiracMisner}

The aim of this appendix is to show that the regularity conditions \eqref{eq:DiracMisner} automatically imply the absence of Dirac-Misner singularities. The latter arise as a consequence of the fact that the 1-forms $\chi$ and $\ii \breve \omega$, given respectively in \eqref{eq:chi} and \eqref{eq:breve_omega}, are not well defined along the $x^3$-axis, where the coordinate $\phi$ is not defined. It is convenient to distinguish the following three regions: \textbf{I} ($x^3>\delta/2$), \textbf{II} ($-\delta/2<x^3<\delta/2$) and \textbf{III} ($x^3<-\delta/2$). Our analysis focuses on the saddles studied in \ref{sec:gen_saddles}, which satisfy ${\breve \omega}_N=-{\breve \omega}_S$. Hence, 1-form $\ii \breve \omega$ is well defined in \textbf{I} and \textbf{III}, but not in \textbf{II}. In turn, $\chi$ is ill-defined in the entire $x^3$-axis. This can be fixed by a coordinate change of $\tau$ and $\psi$. However, since the coordinate change is different in each region, one has to check that they are compatible in the overlaps. We shall see that this is automatically guaranteed once \eqref{eq:DiracMisner} is assumed. The coordinate transformations in each of the regions are the following.
\begin{itemize}
\item \textbf{Region I ($x^3>\delta/2$).}  A coordinate transformation that renders the 1-form $\diff \psi +\chi$ regular is
\begin{equation}
 \left(\tau^I, \psi^I\right)=\left(\tau, \psi+\phi +\ii \left(\Omega_+-\Omega_-\right)\tau\right)\, .
\end{equation}
Explicitly, one has
\begin{equation}
\diff{\psi}+\chi = \diff \psi^I-\ii \left(\Omega_+-\Omega_-\right)\diff\tau^I+ \chi-\diff \phi\, ,
\end{equation}
which is now regular in \textbf{I}. The condition $\beta \Omega_{+}=2\pi \ii$ implies that the coordinates $(\tau^I, \psi^I, \phi)$ have untwisted identifications, inherited from \eqref{eq:twisted_bdry_conditions}:
\begin{equation}
\label{eq:untwisted_id_I}
\left(\tau^I, \psi^I, \phi\right)\sim \left(\tau^I+\beta, \psi^I, \phi\right)\sim \left(\tau^I, \psi^I+4\pi, \phi\right)\sim \left(\tau^I, \psi^I, \phi+2\pi\right).
\end{equation}
\item \textbf {Region II ($-\delta/2<x^3<\delta/2$).} A coordinate transformation rendering regular both $\diff \tau + \ii {\breve \omega}$ and $\diff \psi +\chi$ in this region is 
\begin{equation}
\begin{aligned}
\left(\tau^{II}, \psi^{II}\right)=&\left(\tau-2\ii \breve{\omega}_{N}\phi, \psi+\phi+\frac{h_S-h_N-1}{2\ii {\breve \omega}_N}\tau\right)\\
=&\left(\tau-\frac{\beta}{2\pi} \phi, \psi+\phi+\ii \left(\Omega_+-\Omega_-\right)\tau\right)\, ,
\end{aligned}
\end{equation}
where in the second line we have made use of \eqref{eq:DiracMisner}. Written in this way, it is straightforward to check that $(\tau^{II}, \psi^{II}, \phi)$ satisfy the same untwisted identifications of region \textbf{I}, \eqref{eq:untwisted_id_I}. In the overlap \textbf{I-II}, one has
\begin{equation}
\tau^{II}=\tau^{I}-\frac{\beta}{2\pi}\phi\,, \hspace{1cm} \psi^{II}=\psi^{I} \,,
\end{equation}
which is consistent with the global identifications of the coordinates.

\item \textbf {Region III ($x^3<-\delta/2$).} Finally, the coordinate transformation needed in this region is 
\begin{equation}
 \left(\tau^{III}, \psi^{III}\right)=\left(\tau, \psi-\phi -\ii \left(\Omega_++\Omega_-\right)\tau\right).
\end{equation}
As in region \textbf{I}, the condition $\beta\Omega_{+}=2\pi \ii$ implies that $(\tau^{III}, \psi^{III}, \phi)$ satisfy the untwisted identifications of \eqref{eq:untwisted_id_I}. In the overlap \textbf{II-III}, we have 
\begin{equation}
\tau^{II}=\tau^{III}-\frac{\beta}{2\pi}\phi\,, \hspace{1cm} \psi^{II}=\psi^{III}+2\phi -\frac{4\pi}{\beta}\tau^{III} \,,
\end{equation}
which is again consistent with the global identifications of the coordinates.
\end{itemize}

\bibliography{localizationsugra.bib}
\bibliographystyle{JHEP}

\end{document}